\documentclass[10pt]{book}
\usepackage{sussp2004}
\usepackage{graphicx}
\usepackage{latexsym}
\usepackage{epstopdf}
\begin{document}

\setcounter{section}{1}
\setcounter{page}{1}

\headings{The QCD vacuum and its hadronic excitations}{The QCD vacuum and its hadronic excitations}{W. Weise}{Physik-Department, Technische Universit\"at M\"unchen\\
D-85747 Garching, Germany}

\section{Introduction: QCD - its phases and structures}

These lectures deal with the complex structure of the ground state, or vacuum, of  Quantum Chromo Dynamics (QCD) and its low-energy excitations, the hadrons.
QCD, the theory of quarks, gluons and their interactions, is a self-contained part of the Standard Model of elementary particles. It is a consistent quantum field theory  with a simple and elegant underlying Lagrangian, based entirely on the invariance under a local gauge group, $SU(3)_{color}$. Out of this Lagrangian emerges an enormously rich variety of physical phenomena, structures and phases. Exploring and understanding
these phenomena is undoubtedly one of the most exciting challenges in modern science. 

Fig.1 shows a schematic phase diagramme of QCD for first orientation. At high temperatures, above a critical temperature $T_c$ of about 0.2 GeV, the elementary quark and gluon degrees of freedom are released from their confinement in hadrons. Correlations between these basic constituents are expected still to persist up to quite high temperatures, above which matter presumably exists in the form of a quark-gluon plasma. 

At temperatures below $T_c$ and at low baryon density, matter exists in aggregates of quarks and gluons with their color charges combined to form neutral (color-singlet) objects. This is the domain of low-energy QCD, the physics of the hadronic phase in which mesons, baryons and nuclei reside. In this phase the QCD vacuum has undergone a qualitative change to a ground state chracterised by strong condensates of quark-antiquark 
\centerline {
\includegraphics[width=7cm]{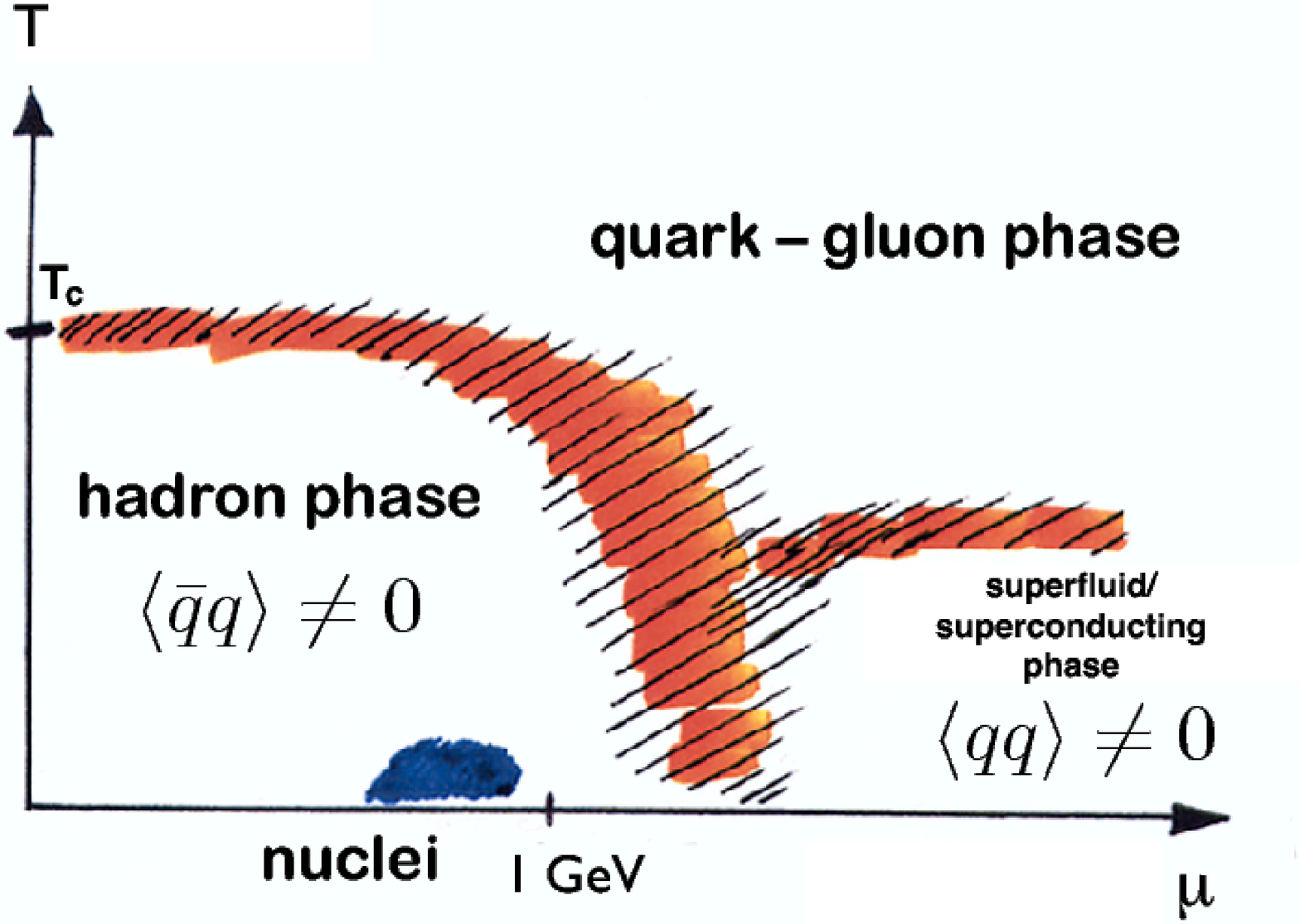}}
\begin{center}
Figure 1: Illustration of the QCD phase diagramme
in the plane of temperature $T$ and baryon chemical potential $\mu$. 
\end{center}
pairs and gluons. In another sector, at very high baryon chemical potentials (i.e. at large quark densities and Fermi momenta), it is expected that Cooper pairing of quarks sets in and induces transitions to a complex pattern of superconducting and superfluid phases.

Major parts of these lectures will be focused on concepts and strategies used to investigate the hadronic sector of the QCD phase diagramme. Basic QCD symmetries and the corresponding conserved currents are used as a guiding principle to construct effective Lagrangians which represent QCD at low energies and momenta. A rapidly advancing approach to deal with non-perturbative QCD is Lattice Gauge Field Theory. Considerable progress is being made solving QCD on a discretised, Euclidean space-time lattice using powerful computers. Applications of effective field theory and some selected examples of lattice QCD results will both be discussed in these notes. Introductory materials have in part been updated from a previous lecture series (Weise 2003).

\section{Basics}

\subsection{QCD primer}

The elementary spin-1/2 particles of QCD, the quarks, come in six species, or flavors, grouped in a field $\psi(x) = (u(x), d(x), s(x), c(x), b(x), t(x))^T$. Each of the $u(x), d(x), ...$ is a four-component Dirac spinor field. Quarks experience all three fundamental interactions of the Standard Model: weak, electromagnetic and strong. Their strong interactions involve $N_c = 3$ "color" charges for each quark. These interactions are mediated by the gluons, the gauge bosons of the underlying gauge group of QCD, $SU(3)_{color}$. 

The QCD Lagrangian (Fritzsch {\it et al} 1973) is
\begin{equation}
{\cal L}_{QCD} = \bar{\psi}\left(i\gamma_\mu{\cal D}^\mu - m\right)\psi -{1\over 4}G_{\mu\nu}^{j}G^{\mu\nu}_{j} \, ,
\end{equation}
with the gauge-covariant derivative
\begin{equation}
{\cal D}^\mu  = \partial_\mu - ig \sum_{j=1}^8 {\lambda_j\over 2} {\cal A}_\mu^{j}(x) \, 
\end{equation}
and the gluon field tensor
\begin{equation}
G_{\mu\nu}^{i}(x) = \partial_\mu {\cal A}_\nu^{i}(x) -  \partial_\nu {\cal A}_\mu^{i}(x) + g f_{ijk} {\cal A}_\mu^{j}(x) {\cal A}_\nu^{k}(x)  \, .
\end{equation}
The $\lambda_j$ are Gell-Mann matrices and the $f_{ijk}$ are the antisymmetric structure constants of the SU(3) Lie algebra. The non-linear three- and four-point couplings of the gluon fields ${\cal A}_\mu^j$ with each other are at the origin of the very special phenomena encountered in QCD and strong interaction physics. 

Apart from the quark masses collected in the mass matrix $m = diag(m_u, m_d, m_s, ... )$ in Eq.(1), there is no primary scale in ${\cal L}_{QCD}$. Renormalisation of the quantum field theory introduces such a scale, to the effect that the "bare" coupling constant $g$ that enters Eqs.(2,3) turns into a "running" coupling, $g(\mu)$, depending on the scale $\mu$ at which it is probed. It is common to introduce the QCD coupling strength as $\alpha_s(\mu) = g^2(\mu)/4\pi$, by analogy with the fine-structure constant in QED. Leading-order perturbative QCD gives
\begin{equation}
\alpha_s(\mu) = {4\pi \over \beta_0\,\ln (\mu^2/\Lambda^2)}\, ,
\end {equation}
with $\beta_0 = 11- 2N_f/3$ where $N_f$ is the number of quark flavors. The QCD scale parameter $\Lambda$ is determined empirically ($\Lambda \simeq 0.2$ GeV for $N_f = 4$). The fact that $\alpha_s$ decreases with increasing $\mu$ leads to a property known as "asymptotic freedom", the domain $\mu \gg 1$ GeV in which QCD can indeed be treated as a perturbative theory of quarks and gluons. The theoretical discovery of asymptotic freedom was honored with the 2004 Nobel Prize in Physics (Gross and Wilczek 1973, Politzer 1973). At the scale of the Z-boson mass, $\alpha_s(M_Z) \simeq 0.12$ (see Particle Data Group 2004). So while $\alpha_s$ is small at large $\mu$,
it is of order one at $\mu < 1$ GeV. At low energies and momenta, an expansion in powers of $\alpha_s$ is therefore no longer justified: we are entering the region commonly referred to as non-perturbative QCD.  
 
The quarks are classified as "light" or "heavy" depending on their entries in the mass matrix $m$ of Eq.(1). These masses are "running" as well: they depend on the scale $\mu$ at which they are determined. The masses of the lightest ($u$ and $d$) quarks, 
\begin{equation}
m_{u,d} < 10\, MeV
\end{equation}
(estimated at a renormalisation  scale $\mu \simeq$ 1 GeV) are very small compared to typical hadron masses of order 1 GeV, such as those of the $\rho$ meson or the nucleon. The strange quark mass, 
\begin{equation}
m_s \simeq (100 - 150)\, MeV
\end{equation} 
is an order of magnitude larger than $m_{u,d}$ but still counted as "small" on hadronic scales. The charm quark mass
$m_c \simeq (1.1 - 1.4)$ GeV takes an intermediate position while the $b$ and $t$ quarks ($m_b \simeq (4.1 - 4.4)$ GeV, \,\,$m_t = (174\pm 5)$ GeV) fall into the "heavy" category. These different quark masses set a hierarchy of scales, each of which is governed by distinct physics phenomena.

\subsection{Concepts and strategies}

There exist two limiting situations in which QCD is accessible with "controlled" approximations. 
At momentum scales exceeding several GeV (corresponding to short distances, r < 0.1 fm), QCD is a theory of weakly interacting quarks and gluons (Perturbative QCD). At low momentum scales considerably smaller than 1 GeV (corresponding to long distances, r > 1 fm), QCD is characterised by confinement and a non-trivial vacuum (ground state) with strong condensates of quarks and gluons.  
Confinement implies the spontaneous breaking of a symmetry which is exact in the limit of massless quarks: chiral symmetry. Spontaneous chiral symmetry breaking in turn implies the existence of pseudoscalar Goldstone bosons. For two flavors ($N_f = 2$) they are identified with the isotriplet of pions ($\pi^+, \pi^0, \pi^-$). For $N_f = 3$, with inclusion of the strange quark, this is generalised to the pseudoscalar meson octet. Low-energy QCD is thus realised as an Effective Field Theory (EFT) in which these Goldstone bosons are the active, light degrees of freedom. 

Much of the interesting physics lies also between these extreme short and long distance limits. This is where large-scale computer simulations of QCD on discretised Euclidean space-time volumes (lattices) are progressing steadily (Davies 2004). Fig.2 presents a key result of lattice QCD (Bali 2001): the gluonic flux tube that connects two infinitely heavy color sources (quark and antiquark) fixed at lattice sites with a given distance. The resulting quark-antiquark potential is well approximated by the form
\begin{equation}
V(r) = -{4\alpha_s\over 3r} + \sigma r\,.
\end{equation}
Coulomb-like (perturbative) one-gluon exchange with $\alpha_s \simeq 0.3$ is seen at short distance. At long range, the potential shows the linear rise with increasing distance between the
\centerline {
\includegraphics[width=10cm]{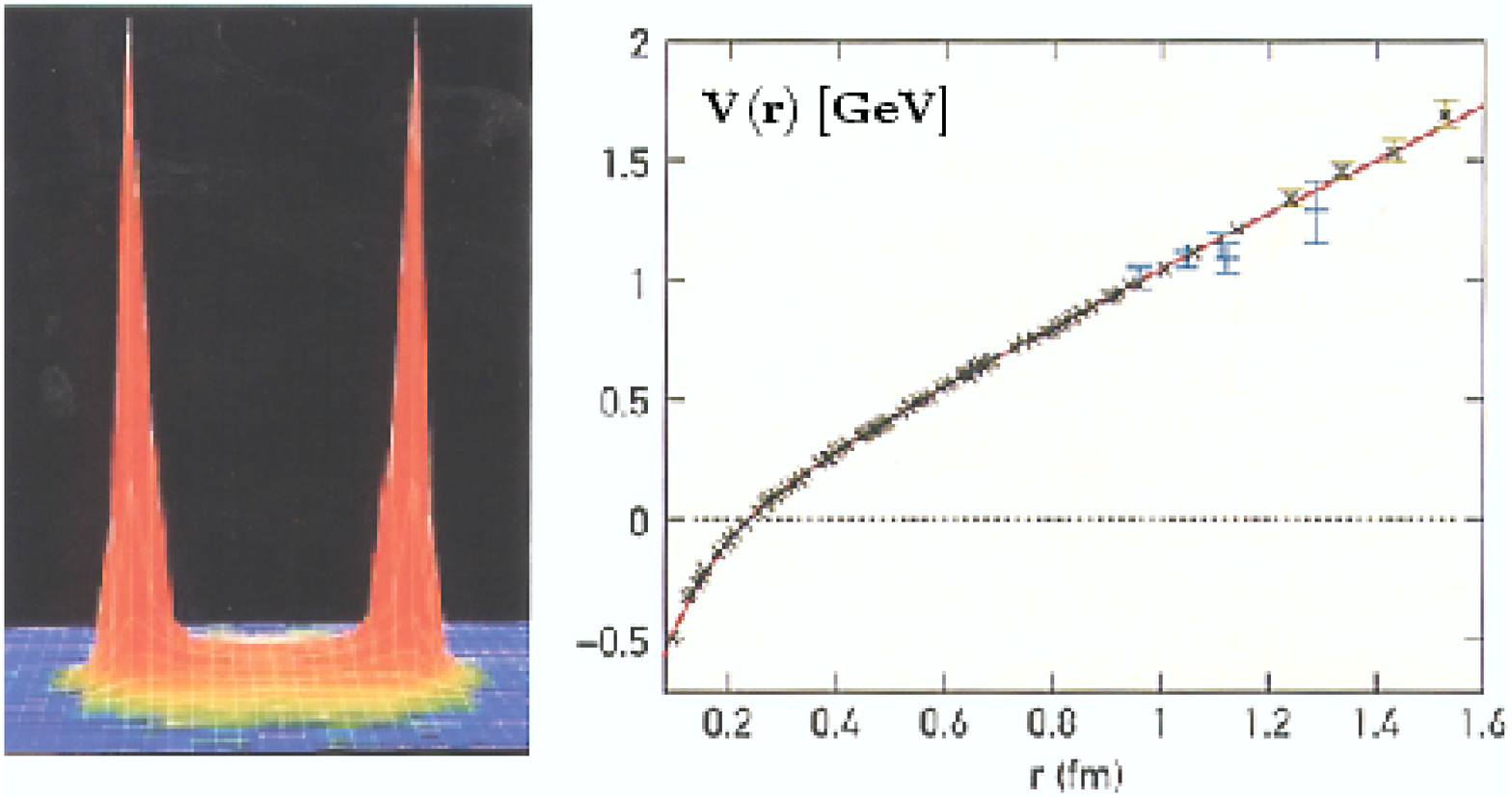}}
Figure 2: Confinement from Lattice QCD (Bali 2001). \\Left: Gluonic flux tube extending between static color sources (infinitely heavy quarks). \\Right: Static quark-antiquark potential in physical units. \\
static color charges, characteristic of confinement, parametrised by a string tension $\sigma \simeq1$ GeV/fm. Confining potentials of the form (7) have been used successfully in accurate decriptions of bottomonium spectroscopy, solving the Schr\"odinger equation for $b\bar{b}$ bound states with inclusion of spin-dependent interactions (Bali {\it et al.}, 1997).

Confinement has a relatively simple interpretation for heavy quarks and the "string" of (static) gluonic field strength that holds them together, expressed in terms of a static potential. When light quarks are involved, the situation is different. Color singlet quark-antiquark pairs pop out of the vacuum as the gluon fields propagate over larger distances. Light quarks are fast movers: they do not act as static sources. The potential picture is not applicable.  The common features of the confinement phenomenon can nevertheless be phrased as follows: non-linear gluon dynamics in QCD does not permit the propagation of colored objects over distances of more than a fraction of a Fermi. Beyond the one-Fermi scale, the only remaining relevant degrees of freedom are color-singlet composites (quasiparticles) of quarks, antiquarks and gluons.

\subsection{Scales at work: the hierarchy of quark masses} 

The quark masses are the only parameters that set primary scales in QCD. Their classification into sectors of "light" and "heavy" quarks determines the very different physics phenomena associated with them. The heavy quarks (i.e. the $t$, $b$ and - within limits - the $c$ quarks) offer a natural "small parameter" in terms of their reciprocal masses. Non-relativistic approximations (expansions of observables in powers of $m_{t,b,c}^{-1}$) tend to work increasingly well with increasing quark mass. This is the area sometimes referred to under the name of Non-Relativistic QCD.

The sector of the light quarks (i.e. the $u$, $d$ quarks and - to some extent - the $s$ quarks) is governed by quite different principles and rules. Evidently, the quark masses themselves are now "small parameters", to be compared with a characteristic "large" scale of dynamical origin. This large scale is related to the order parameter of spontaneously broken chiral symmetry, commonly identified with $4\pi f_\pi \sim 1$ GeV, where $f_\pi \simeq 0.09$ GeV is the pion decay constant to be specified later. The square of $f_\pi$ is, in turn, proportional to the chiral (or quark) condensate $\langle\bar{q}q\rangle$, a key quantity featuring the non-trivial structure of the QCD vacuum. The low-mass hadrons are quasiparticle excitations of this condensed ground state. There is a characteristic mass gap of about 1 GeV which separates the QCD vacuum from almost all of its excitations, with the exception of the pseudoscalar meson octet of pions, kaons and the eta meson. This mass gap is again comparable to $4\pi f_\pi$, the scale associated with spontaneous chiral symmetry breaking in QCD. 

Low-energy QCD is the physics of systems of light quarks at energy and momentum scales smaller than the 1 GeV mass gap observed in the hadron spectrum. The relevant expansion parameter in this domain is $Q/4\pi f_\pi$, where $Q$ stand generically for low energy, momentum or pion mass.

In this context it is instructive to explore the spectroscopic pattern of pseudoscalar and vector mesons,
starting from heavy-light quark-antiquark pairs in $^1S_0$ and $^3S_1$ states and following those states downward in the mass of the quark. This is illustrated in Fig.3 where we show the masses of mesons composed of a $b$, $c$, $s$ or $u$ quark with an anti-$d$-quark attached. Bare quark masses are subtracted from the meson masses in this plot in order to directly demonstrate the evolution from perturbative hyperfine splitting in the heavy systems to the non-perturbative mass gap in the light ones. In the $\bar{B}$ and $\bar{B}^*$ mesons, the $\bar{d}$ quark is tightly bound to the heavy $b$ quark at small average distance, within the range where perturbative QCD is applicable. The spin-spin interaction is well approximated by perturbative one-gluon exchange, resulting in a small hyperfine splitting. Moving downward in mass to the $D$ and $D^*$ systems, with the $b$ quark replaced by a $c$ quark, the hyperfine splitting increases but remains perturbative in magnitude. As this pattern evolves further into the light-quark sector, it undergoes a qualitative change via the large mass difference of $\bar{K}$ and $\bar{K}^*$  to the non-perturbative mass gap in the $\pi - \rho$ system, reflecting the Goldstone boson nature of the pion.\\

\centerline {
\includegraphics[width=8cm]{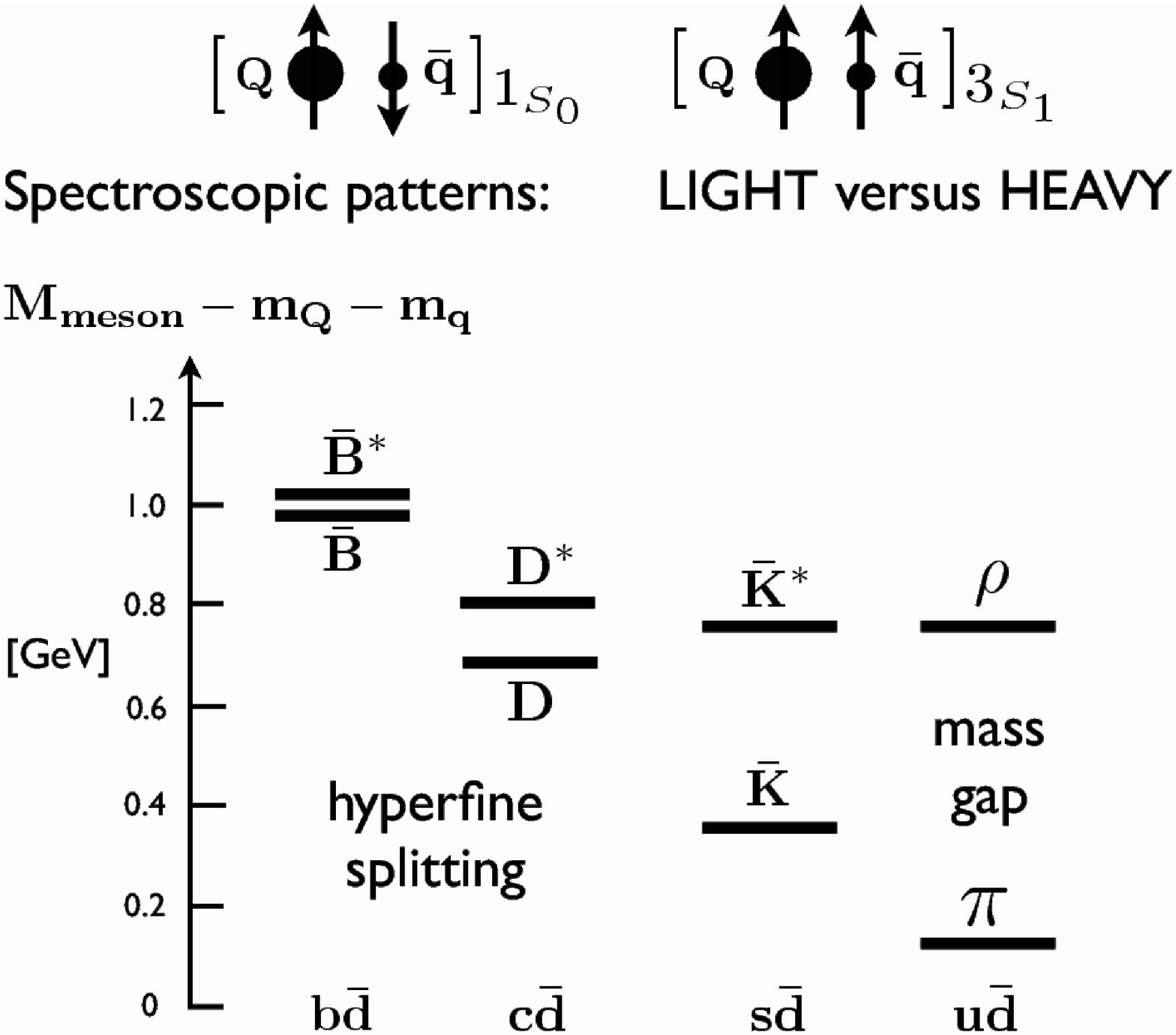}}
\begin{center}
Figure 3: Evolution of the splitting between spin singlet (lower) and triplet (upper) quark-antiquark states
(the pseudoscalar ($J^\pi = 0^-$) and vector ($J^\pi = 1^-$) mesons) when varying the mass of one of the quarks. The bare quark masses are subtracted from the physical meson masses for convenience of demonstration. 
\end{center}

\section{Low-energy QCD}

In the hadronic, low-energy phase of QCD, the active degrees of freedom are not elementary quarks and gluons but mesons and baryons. To illustrate the situation, consider the canonical partition function expressed in terms of the QCD Hamiltonian $H$:
\begin{eqnarray}
{\cal Z} = Tr \,\exp (-H/T) = \sum_n \langle n | e^{-E_n /T} | n \rangle 
\nonumber
\end{eqnarray}
with $(H - E_n) | n \rangle = 0$. Confinement implies that the eigenstates $| n
\rangle$ of $H$ are (colour-singlet) hadrons at $T < T_c$, below the critical temperature for deconfinement. The low-temperature
physics is then determined by the states of lowest mass in the spectrum $\{ E_n
\}$. The very lowest states are the pseudoscalar mesons. All other states are separated
by the mass gap mentioned earlier. This separation of scales is the key for approaching low-energy QCD with the systematic theoretical tool of Chiral Effective Field Theory, to be developed in the
next chapter. The guiding principle for this is the approximate chiral symmetry of QCD.

\subsection{Chiral symmetry}

Consider QCD in the limit of massless quarks, setting $m=0$ in Eq.(1). In this
limit, the QCD Lagrangian has a global symmetry related to the conserved right-
or left-handedness (chirality) of zero mass spin $1/2$ particles. We
concentrate here first on the $N_f = 2$ sector of the two lightest ($u$- and $d$-)
quarks with $\psi(x) = (u(x), d(x))^T$. Introducing right- and left-handed quark fields,
\begin{equation}
\psi_{R,L} = \frac{1}{2} (1 \pm \gamma_5) \psi ,
\end{equation}
we observe that separate global unitary transformations
\begin{equation}
\psi_R \to \exp [i \theta^a_R \frac{\tau_a}{2}] \, \psi_R, \hspace{1,5cm} \psi_L
\to \exp [i \theta^a_L \frac{\tau_a}{2}] \, \psi_L ,
\end{equation}
with $\tau_a\,(a= 1,2,3)$ the generators of the $SU(2)$ flavor group, leave ${\cal L}_{QCD}$
invariant in the limit $m \to 0\,$: the right- and left-handed components of the
massless quark fields do not mix. This is the chiral $SU(2)_R \times
SU(2)_L$ symmetry of QCD. It implies six conserved Noether currents,
$J^{\mu}_{R, a} = \bar{\psi}_R \gamma^{\mu} \frac{\tau_a}{2} \psi_R$ and
$J^{\mu}_{L, a} = \bar{\psi}_L \gamma^{\mu} \frac{\tau_a}{2} \psi_L$, with
$\partial_{\mu} J^{\mu}_R = \partial_{\mu} J^{\mu}_L = 0 $. It is common to
introduce the vector current
\begin{equation}
V^{\mu}_a = J^{\mu}_{R,a} + J^{\mu}_{L,a} = \bar{\psi} \gamma^{\mu}
\frac{\tau_a}{2} \psi ,
\end{equation}
and the axial current,
\begin{equation}
A^{\mu}_a (x) = J_{R,a} - J_{L,a} = \bar{\psi} \gamma^{\mu} \gamma_5 \frac{\tau_a}{2} \psi .
\end{equation}
Their corresponding charges,
\begin{equation}
Q^V_a = \int d^3 x 
~\psi^{\dagger} (x) \frac{\tau_a}{2} \psi (x), \hspace{1,5cm}
Q^A_a = \int d^3 x 
~\psi^{\dagger} (x) \gamma_5 \frac{\tau_a}{2} \psi (x) ,
\end{equation}
are, likewise, generators of $SU(2) \times SU(2)$.

To the extent that the strange quark mass $m_s$ can also be considered "small",
it makes sense to generalize the chiral symmetry to $N_f = 3$. The three Pauli
matrices $\tau_a$ are then replaced 
by the eight Gell-Mann matrices $\lambda_a$ of $SU(3)$.

\subsection{Spontaneous symmetry breaking}

There is evidence from hadron spectroscopy that the chiral  
$SU(2) \times SU(2)$ symmetry 
of the QCD Lagrangian (1) with $m = 0$ is
spontaneously broken: for dynamical reasons of non-perturbative
origin, the ground state (vacuum) of QCD
is symmetric only under the subgroup $SU(2)_V$ generated by the vector charges
$Q^V$. This is the well-known isospin symmetry or, correspondingly, the
"eightfold way" when extended to $N_f = 3$.

If the ground state of QCD were symmetric under chiral $SU(2) \times SU(2)$,
both vector and axial charge operators (12) would annihilate the vacuum: $Q^V_a |0 \rangle
= Q^A_a |0 \rangle = 0$. This is the Wigner-Weyl realisation of chiral symmetry with
a "trivial" vacuum. It would imply the systematic appearance of parity doublets
in the hadron spectrum. 

Consider for example the correlation functions of vector and axial vector currents:
\begin{equation}
\Pi_V^{\mu\nu}(q) = i \int d^4x\,e^{iq\cdot x}  \langle 0|{\cal T}[V^\mu(x)V^\nu(0)]|0\rangle \, , 
\end{equation}
\begin{equation}
\Pi_A^{\mu\nu}(q) = i \int d^4x\,e^{iq\cdot x}  \langle 0|{\cal T}[A^\mu(x)A^\nu(0)]|0\rangle \, , 
\end{equation}
where ${\cal T}$ denotes time ordering and $\Pi^{\mu\nu}(q) = (q^\mu q^\nu - q^2 g^{\mu\nu})\Pi(q^2)$.
If chiral symmetry were in its (trivial) Wigner-Weyl realisation, these two correlation functions should be identical: $\Pi_V = \Pi_A$. Consequently, their spectral distributions 
\begin{equation}
\eta_{V,A}(s) = 4\pi\,Im\,\Pi_{V,A}(s = q^2)
\end{equation}
which include the vector $(J^{\pi} = 1^-)$ and axial vector $(J^{\pi} = 1^+)$ mesonic
excitations, should also be identical. This degeneracy is not observed in nature. The 
$\rho$ meson mass ($m_{\rho} \simeq 0.77$ GeV) is well separated from
that of the $a_1$ meson ($m_{a_1} \simeq 1.23$ GeV), as can be seen (Fig.4) in the resonance
spectra measured in $\tau$ decays to the relevant channels. Likewise, the light
pseudoscalar $(J^{\pi} = 0^-)$ mesons have masses much lower than the lightest
scalar $(J^{\pi} = 0^+)$ mesons.
\\ \\
\begin{minipage}[t]{6cm}
\centerline {
\includegraphics[width=5.9cm]{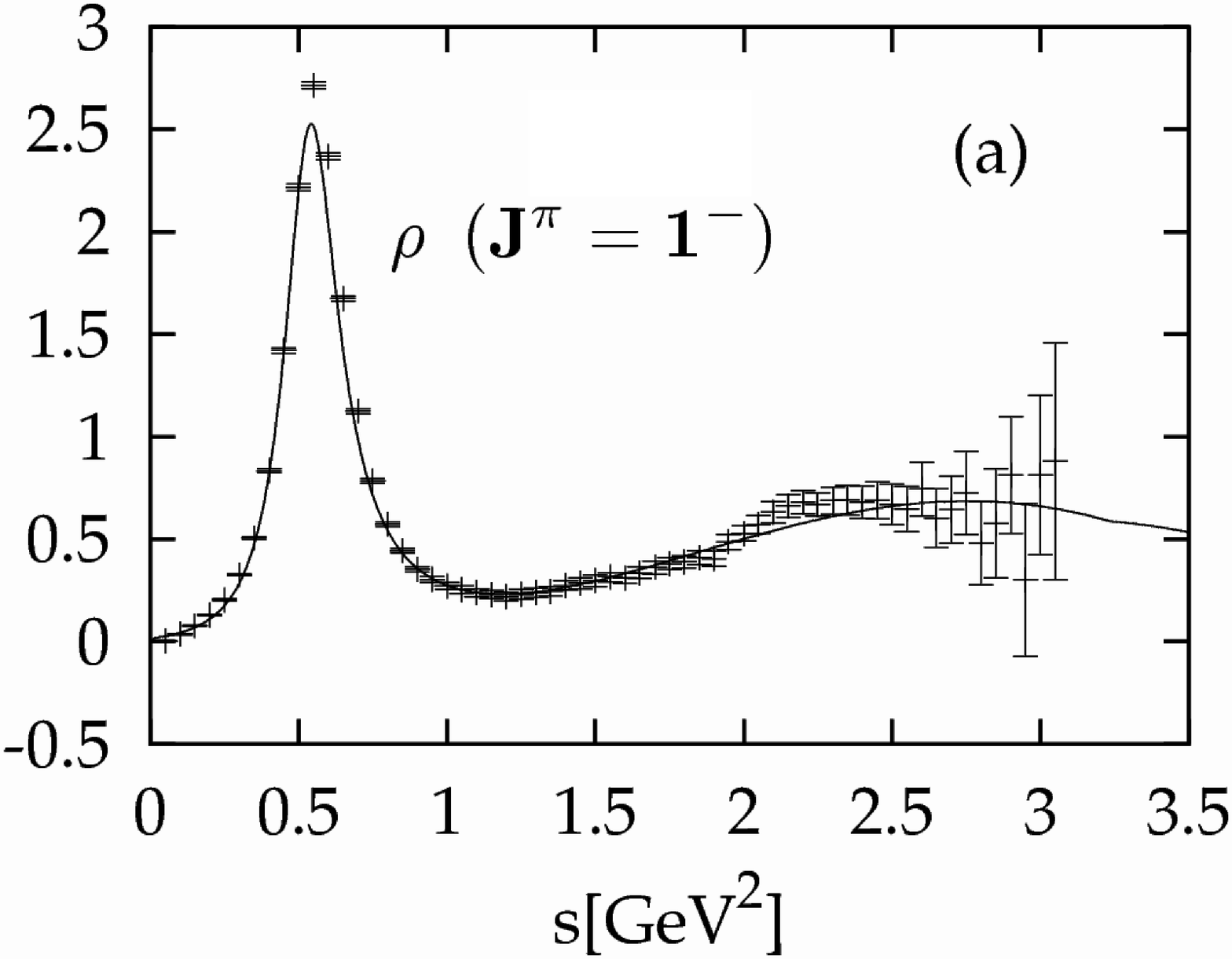}}
\end{minipage}
\hspace{\fill}
\begin{minipage}[t]{6.2cm}
\centerline {
\includegraphics[width=5.95cm]{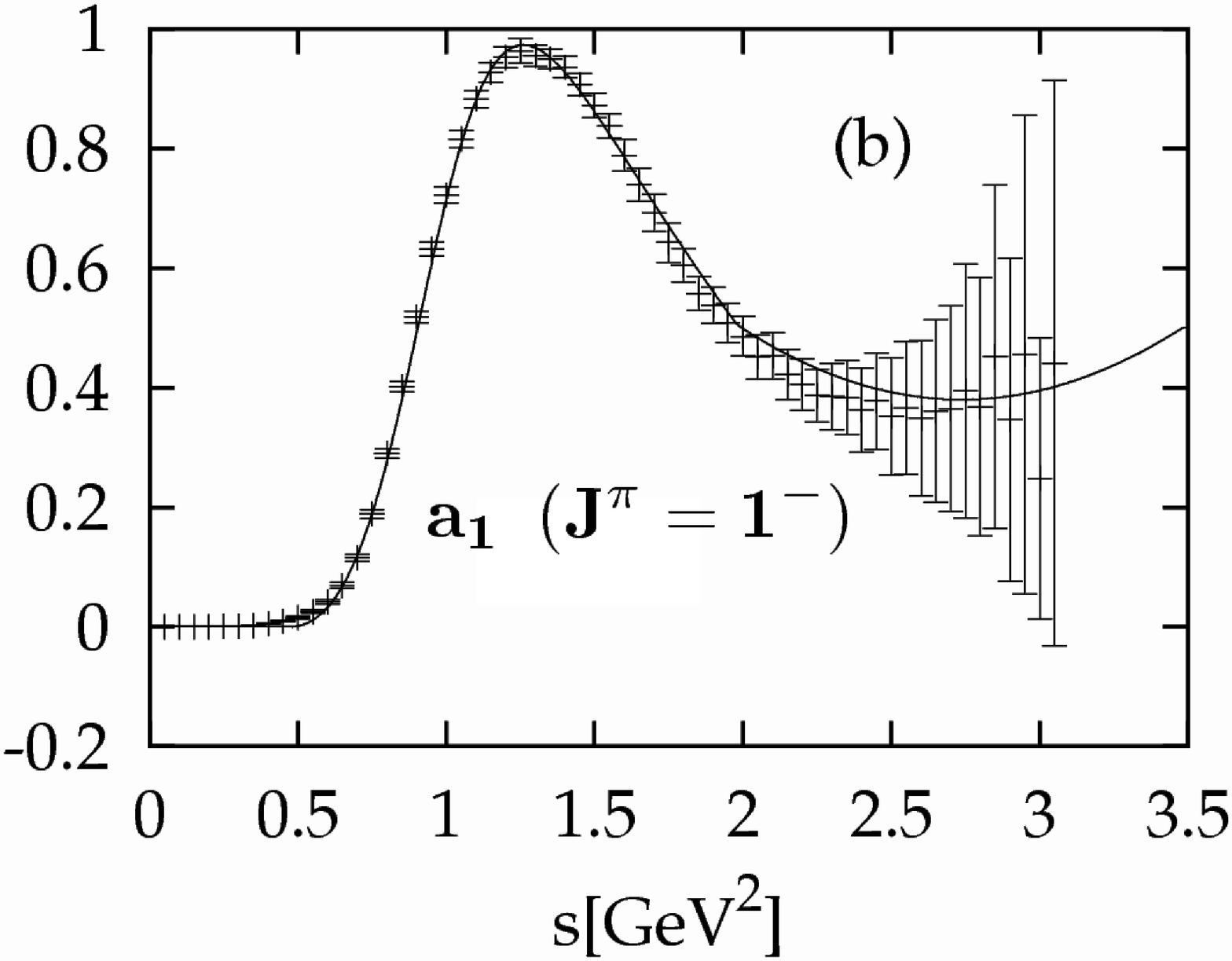}}
\end{minipage}
Figure 4: Spectral distributions $\eta(s)$ of vector (a) and axial vector (b) mesons from $\tau$ decays (Barate {\it et al.} 1998, Ackerstaff {\it et al.} 1999). The curves are results of a QCD sum rule analysis (Marco and Weise 2000). 
\\

One must conclude that $Q^A_a |0 \rangle \neq 0$, that is, chiral symmetry is
spontaneously broken down to isospin: $SU(2)_R \times SU(2)_L \to
SU(2)_V$. This is the Nambu-Goldstone realisation of chiral symmetry.

A spontaneously broken global symmetry implies the existence of (massless)
Goldstone bosons. If $Q^A_a | 0 \rangle \neq 0$, there must be a physical state
generated by the axial charge, $|\phi_a \rangle = Q^A_a | 0 \rangle$, which is
energetically degenerate with the vacuum. Let $H_0$ be the QCD Hamiltonian
(with massless quarks) which commutes with the axial charge. Setting the ground
state energy equal to zero for convenience, we have 
$H_0 |\phi_a \rangle = Q^A_a
H_0 | 0 \rangle = 0$. Evidently $| \phi_a \rangle $ represents three massless
pseudoscalar bosons (for $N_f = 2$). They are identified with the isotriplet of pions.

When the spontaneous symmetry breaking scenario is extended to chiral $SU(3)_R
\times SU(3)_L \to SU(3)_V$, the corresponding eight Goldstone bosons are
identified with the members of the lightest pseudoscalar octet: pions, kaons,
antikaons and the $\eta$ meson. The $\eta '$ meson, on the other hand, falls
out of this scheme. The large $\eta'$ mass reflects the axial
$U(1)_A$ anomaly in QCD. Without this anomaly, QCD with $N_f = 3$ massless
quarks would actually have a chiral $U(3)_R \times U(3)_L$ symmetry, and its
spontaneous breakdown would lead to nine rather than eight pseudoscalar
Goldstone bosons. The axial anomaly removes the $U(1)_A$ symmetry, keeping
$SU(3)_R \times SU(3)_L \times U(1)_V$ intact which is then spontaneously
broken down to $SU(3)_V \times U(1)_V$. The remaining $SU(3)$ flavour symmetry
is accompanied by the conserved baryon number which generates
$U(1)_V$.

\subsection{The chiral condensate}

Spontaneous chiral symmetry breaking goes in parallel with a qualitative
re-arrangement of the vacuum, an entirely non-perturbative phenomenon. The
ground state is now populated by scalar quark-antiquark pairs. The
corresponding ground state expectation value $\langle 0 | \bar{\psi} \psi | 0
\rangle$ is called the chiral condensate (also referred to under the name
"quark condensate"). We frequently use the notation
\begin{equation}
\langle \bar{\psi} \psi \rangle = \langle \bar{u} u \rangle + \langle \bar{d} d
\rangle
\end{equation}
(for two flavours, with $\langle \bar{s} s \rangle$ added when going to $N_f =
3$). The precise definition of the chiral condensate is:
\begin{equation}
\langle \bar{\psi} \psi \rangle = -i Tr \lim_{y \to x^+} S_F (x,y)
\end{equation}
with the full quark propagator, $S_F (x,y) = -i \langle 0 | {\cal T} \psi (x)
\bar{\psi} (y) | 0 \rangle$. We recall Wick's theorem which states that the time-ordered 
product ${\cal T} \psi (x)
\bar{\psi} (y)$ reduces to the normal product $ :\psi (x) \bar{\psi} (y):$ plus
the contraction of the two field operators. When considering the perturbative
quark propagator, $S^{(0)}_F (x,y)$, the time-ordered product is taken with
respect to a trivial vacuum for which the expectation value of $ : \bar{\psi}
\psi :$ vanishes. Long-range, non-perturbative physics is then at the origin of
a non-vanishing $\langle : \bar{\psi} \psi  : \rangle$ in Eq.(17). (In order
to ensure that no perturbative pieces are left in $\langle \bar{\psi} \psi
\rangle$ for the case of non-zero quark masses, one should actually make the
replacement $S_F \to S_F - S_F^{(0)}$ in (17)).

Let us establish the connection between spontaneous chiral symmetry breaking
and the non-vanishing chiral condensate in a more formal way. Introduce the
pseudoscalar quantity $P_a (x) = \bar{\psi} (x) \gamma_5 \tau_a \psi (x)$ and
derive the (equal-time) commutator relation
\begin{equation}
[ Q^A_a, P_b] = - \delta_{ab}\, \bar{\psi} \psi
\end {equation}
which involves the axial charge $Q^A_a$ of Eq.(12). Taking the ground state
expectation value on both sides of Eq.(18), we see that $Q^A_a | 0 \rangle \neq
0$ is indeed consistent with $\langle \bar{\psi} \psi \rangle \neq 0$.

At this point a brief digression is useful in order to demonstrate spontaneous chiral symmetry breaking and its restoration at high temperatures as seen in computer simulations of QCD thermodynamics. In lattice QCD the path integral defining the partition function ${\cal Z}$ is regularised on a four-dimensional space-time grid with lattice constant $a$. Given $N_\sigma$ lattice points along each space direction and $N_\tau$ points along the (Euclidean) time axis with $\tau = it$, volume and temperature are specified as
$V= (N_\sigma\, a)^3$ and $T=(N_\tau\, a)^{-1}$. The chiral condensate $\langle \bar{\psi}\psi\rangle_T$ at finite temperature is derived starting from the pressure
\begin{equation}
P=T{\partial\over\partial V}\ln{\cal Z}
\end{equation}
by taking the derivative with respect to the quark mass:
\begin{equation}
\langle\bar{\psi}\psi\rangle_T \sim {\partial P(T,V)\over\partial m_q}\, .
\end{equation}

An example of a lattice QCD result for $\langle \bar{\psi}\psi\rangle_T$ is given in Fig.5. The magnitude of the quark condensate decreases from its $T = 0$ value to zero beyond a critical temperature $T_c$. At $T > T_c$ chiral symmetry is restored in its Wigner-Weyl realisation. At $T<T_c$ the symmetry is spontaneously broken in the low-temperature (Nambu-Goldstone) phase. When extrapolated to
zero quark masses, one finds a chiral 1st-order transition with critical temperatures depending on the number of flavours (Karsch {\it et al.} 2001):
$T_c = (173\pm 8)$ MeV for $N_f = 2$ and $T_c = (154\pm 8)$ MeV for $N_f = 3$. The physically relevant situation, with chiral symmetry explicitly broken by finite quark masses $m_{u,d}\sim 5$ MeV and $m_s\sim 120$ MeV, is expected to be not a phase transition but a crossover with a rapid but
continuous change of $\langle \bar{\psi}\psi\rangle_T$, as in Fig.5.
\centerline {
\includegraphics[width=7cm]{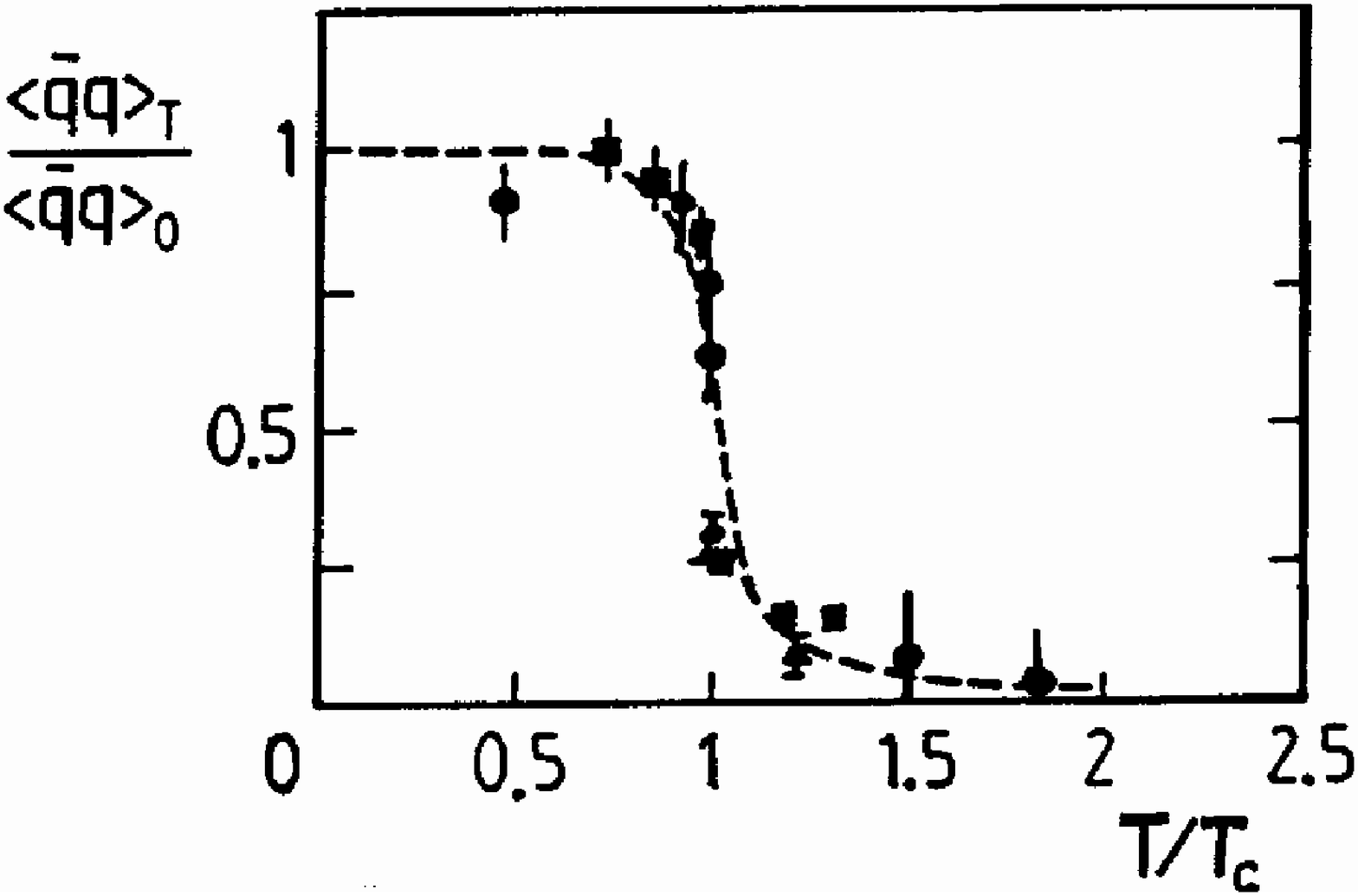}}
\begin{center}
Figure 5: Temperature dependence of the chiral (quark) condensate from lattice QCD (Boyd {\it et al.} 1995). 
\end{center}

Spontaneous symmetry breaking is a widespread phenomenon, and analogues exist
in various other areas of physics. Fig.5 is in fact reminiscent of the picture for the magnetisation of a ferromagnet. The basic
Hamiltonian of this spin system is invariant under rotations in space. However,
its low temperature phase is characterised by a non-vanishing magnetisation
which points into a definite direction in space. Rotational symmetry is
spontaneously broken. This is a non-perturbative, collective phenomenon. Many
spins cooperate to form the macroscopically magnetised material. Slow
variations in the direction of the magnetisation produce a collective
low-frequency, long wavelength motion of the spins. This spin wave (magnon) is
the Goldstone boson of spontaneously broken rotational symmetry. In our case of
spontaneous $\it{chiral}$ symmetry breaking, the analogue of the non-vanishing
magnetisation is the chiral condensate; the analogue of the spin wave is the
pion.

\subsection{PCAC}

Let $| \pi_a (p) \rangle $ be the state vectors of the Goldstone bosons 
associated
with the spontaneous breakdown of chiral symmetry. Their four-momenta are
denoted $p^{\mu} = (E_p, \vec{p}~)$, and we choose the standard normalisation $
\langle \pi_a (p) | \pi_b (p') \rangle = 2 E_p \delta_{ab} (2 \pi)^3 \delta^3
(\vec{p} - \vec{p} \, ')$. Goldstone's theorem, briefly sketched in section 3.2,
also implies non-vanishing matrix elements of the axial current (11) which
connect $| \pi_a (p) \rangle$ with the vacuum:
\begin{equation}
\langle 0 | A^{\mu}_a (x) | \pi_b (p) \rangle = i p^{\mu} f \,\delta_{ab}\, e^{-ip
\cdot x} .
\end{equation}
The constant $f$ is called the pion decay constant (taken here in the chiral
limit, i.~e. for vanishing quark mass). Its physical value (Particle Data Group 2004)
\begin{equation}
f_\pi = (92.4 \pm 0.3) \, MeV
\end{equation}
is determined from the decay $\pi^+ \rightarrow \mu^+ \nu_\mu +  \mu^+ \nu_\mu \gamma$. The
difference between $f$ and $f_{\pi}$ is a correction linear in the quark mass $m_q$.

Non-zero quark masses $m_{u,d}$ shift the mass of the Goldstone boson from zero
to the observed value of the physical pion mass, $m_{\pi}$. The relationship
between $m_{\pi}$ and the $u$ and $d$ quark masses is derived as follows. We
start by observing that the divergence of the axial current (11) is
\begin{equation}
\partial_{\mu} A^{\mu}_a = i \bar{\psi} \{ m, \, \frac{\tau_a}{2} \} \gamma_5
\psi ,
\end{equation}
where $m$ is the quark mass matrix and $\{ , \}$ denotes the anti-commutator. 
This is the microscopic basis for PCAC, the
Partially Conserved Axial Current (exactly conserved in the limit $m \to 0$)
which plays a key role in the weak interactions of hadrons and the low-energy
dynamics involving pions. Consider for example the $a = 1$ component of the 
axial current:
\begin{equation}
\partial_{\mu} A^{\mu}_1 = (m_u + m_d)\, \bar{\psi} i \gamma_5 \frac{\tau_1}{2}
\psi ,
\end{equation}
and combine this with Eq.(18) to obtain
\begin{equation}
\langle 0 | [Q^A_1, \, \partial_{\mu} A^{\mu}_1 ] | 0 \rangle = - \frac{i}{2}
(m_u + m_d) \langle \bar{u} u + \bar{d} d \rangle .
\end{equation}
Now insert a complete set of (pseudoscalar) states in the commutator 
on the left. Assume, in the
spirit of PCAC, that this spectrum of states is saturated by the pion. Then use
Eq.(21) to evaluate $\langle 0 | Q^A | \pi \rangle$ and $\langle 0 |
\partial_{\mu} A^{\mu} | \pi \rangle$ at time $t=0$, with $E_p = m_{\pi}$ at
$\vec{p} = 0$, and arrive at the Gell-Mann, Oakes, Renner (GOR) relation
(Gell-Mann {\it et al.} 1968):
\begin{equation}
m^2_{\pi}\,f_{\pi}^2 = - (m_u + m_d) \langle \bar{q} q \rangle + 
{\cal O}(m^2_{u,d}) .
\end{equation}
We have set $\langle \bar{q} q \rangle \equiv \langle \bar{u} u \rangle \simeq
\langle \bar{d} d \rangle$ making use of isospin symmetry which is valid to a
good approximation. Neglecting terms of order $m^2_{u,d}$
(identifying $f = f_{\pi} = 92.4$ MeV to this order) and inserting
$m_u + m_d \simeq 13$ MeV (Pich and Prades 2000) at a renormalisation 
scale of order 1 GeV, one obtains
\begin{equation}
\langle \bar{q} q \rangle \simeq - (0.23 \pm 0.03 \, GeV)^3 \simeq -1.6  \, fm^{-3} .
\end{equation}
This condensate (or correspondingly, the pion decay constant $f_{\pi}$) is a
measure of spontaneous chiral symmetry breaking. The non-zero pion mass, on the
other hand, reflects the explicit symmetry breaking by the small quark masses,
with $m^2_{\pi} \sim m_q$. It is important to note that $m_q$ and $\langle
\bar{q} q \rangle$ are both scale dependent quantities. Only their product
$m_q \langle \bar{q} q \rangle$ is scale independent, i.e. invariant under the renormalisation
group.

\subsection{The mass gap and spontaneous chiral symmetry breaking}

The appearance of the mass gap $\Gamma \sim 1$ GeV in the hadron spectrum 
is thought to be closely linked to the presence
of the chiral condensate $\langle \bar{\psi} \psi \rangle$ in the QCD ground
state. For example, Ioffe's formula (Ioffe 1981), based on QCD sum rules, connects
the nucleon mass $M_N$ directly with  $\langle \bar{\psi} \psi \rangle$  in
leading order:
\begin{equation}
M_N = - \frac{4 \pi^2}{\Lambda^2_B} \langle \bar{\psi} \psi \rangle + ...~~ ,
\end{equation}
where $\Lambda_B \sim 1$ GeV is an auxiliary scale (the Borel scale) 
which separates
"short" and "long"-distance physics in the QCD sum rule analysis. While this formula is not very accurate and needs to be improved
by including higher order condensates, it nevertheless demonstrates that
spontaneous chiral symmetry breaking plays an essential role in giving the
nucleon its mass.

The condensate  $\langle \bar{\psi} \psi \rangle$ is encoded in the pion decay
constant $f_{\pi}$ through the GOR relation (26). In the chiral limit $(m_q \to
0)$, this $f_{\pi}$ is the only quantity which can serve to define a mass scale
("transmuted" from the QCD scale $\Lambda$ in Eq.(4) through non-perturbative
dynamics). It is common to introduce $ 4 \pi
f_{\pi} \sim 1$ GeV as the scale characteristic of spontaneous chiral symmetry
breaking. This scale is then roughly identified with the spectral
gap $\Gamma$. 

As another typical example of how the chiral gap translates into hadron masses,
consider the $\rho$ and $a_1$ mesons. Finite-energy sum rules for
vector and axial vector current-current correlation
functions, when combined with the algebra of these currents as implied by the
chiral $SU(2) \times SU(2)$ group, do in fact connect the $\rho$ and $a_1$
masses directly with the chiral gap (Weinberg 1967, Klingl and Weise 1999, Goltermann and Peris 2000, Marco and Weise 2000):
\begin{equation}
m_{a_1} = \sqrt{2} \,m_{\rho} = 4 \pi f_{\pi} ,
\end{equation}
at least in leading order (that is, in the large $N_c$ limit, and ignoring
decay widths as well as perturbative QCD corrections).

The relations (28,29), while not accurate at a quantitative level, give
important hints. Systems characterized by an energy gap usually exhibit
qualitative changes when exposed to variations of thermodynamic conditions. A
typical example is the temperature dependence of the gap in a
superconductor. Exploring systematic changes of hadronic spectral functions in a dense and hot medium is therefore a key issue in nuclear and hadron physics (Metag 2004).

\section{Effective field theory}

\subsection{Framework and rules}

The scale set by the mass gap $\Gamma \sim 4 \pi f_{\pi}$ offers a
natural separation between "light" and "heavy" (or, correspondingly, "fast" and
"slow") degrees of freedom. The basic idea of an effective field theory is to
introduce the active light particles as collective degrees of freedom, 
while the
heavy particles are frozen and treated as (almost) static sources. The dynamics
is described by an effective Lagrangian which incorporates all relevant
symmetries of the underlying fundamental theory. We now summarise the necessary
steps, first for the pure meson sector (baryon number $B = 0)$ and later for
the $B = 1$ sector. We work mostly with $N_f = 2$ in this chapter and turn to the $N_f = 3$ case later.

a) The elementary quarks and gluons of QCD are replaced by Goldstone
bosons. They are represented by a $2 \times 2$ matrix field $U (x) \in
SU(2)$ which collects the three isospin components $\pi_a (x)$ of 
the Goldstone pion. A
convenient choice of coordinates is
\begin{equation}
U (x) = \exp[i \tau_a \phi_a(x)]~~ ,
\end{equation}
with $\phi_a = \pi_a/f$ where the pion decay constant $f$ in the chiral limit
provides a suitable normalisation. (Other choices, such
as $U = \sqrt{1- \pi^2_a / f^2} + i \tau_a \pi_a /f$, are also common.
Results obtained from the effective theory must be independent of the
coordinates used.)

b) Goldstone bosons interact weakly at low energy. In fact, if $| \pi
\rangle = Q^A | 0 \rangle$ is a massless state with $H | \pi \rangle = 0$, then
a state $| \pi^n \rangle = (Q^A)^n | 0 \rangle$ with $n$ Goldstone bosons is
also massless since the axial charges $Q^A$ all commute with the full
Hamiltonian $H$. Interactions between Goldstone bosons must therefore vanish at
zero momentum and in the chiral limit.

c) The QCD Lagrangian (1) is replaced by an effective Lagrangian which
involves the field $U (x)$ and its derivatives:
\begin{equation}
{\cal L}_{QCD} \to {\cal L}_{eff} (U, \partial U, \partial^2 U, ...) .
\end{equation}
Goldstone bosons can only interact when they carry momentum, so the low-energy
expansion of (31) is an ordering in powers of $\partial_{\mu} U$. Lorentz
invariance permits only even numbers of derivatives. We write
\begin{equation}
{\cal L}_{eff} = {\cal L}^{(2)} + {\cal L}^{(4)} + ...
\end{equation}
and omit an irrelevant constant term. The leading term, called non-linear sigma
model, involves two derivatives:
\begin{equation}
{\cal L}^{(2)} = \frac{f^2}{4} Tr [ \partial_{\mu} U^{\dagger} \partial^{\mu} U
] .
\end{equation}
At fourth order, the terms permitted by symmetries are (apart from an extra
contribution from the QCD anomaly, not included here):
\begin{equation}
{\cal L}^{(4)} = \frac{l_1}{4} ( Tr [\partial_{\mu} U^{\dagger}
\partial^{\mu} U])^2 +
\frac{l_2}{4} Tr [\partial_{\mu} U^{\dagger} \partial_{\nu} U] Tr
[\partial^{\mu} U^{\dagger} \partial^{\nu} U] ,
\end{equation}
and so forth. The constants $l_1, l_2$ (following canonical notations of
Gasser and Leutwyler 1984) must be determined by experiment. To the extent that the
effective Lagrangian includes all terms dictated by the symmetries, the chiral
effective field theory is the low-energy equivalent of the original QCD Lagrangian 
(Weinberg 1979, Leutwyler 1994).

d) The symmetry breaking mass term is small, so that it can be handled
perturbatively, together with the power series in momentum. The leading
contribution introduces a term linear in the quark mass matrix $m$:
\begin{equation}
{\cal L}^{(2)} = \frac{f^2}{4} Tr [\partial_{\mu} U^{\dagger} \partial^{\mu} U]
+ \frac{f^2}{2} B_0 \, Tr [m (U + U^{\dagger})] .
\end{equation}
The fourth order term ${\cal
L}^{(4)}$ also receives symmetry breaking contributions with additional constants
$l_i$.

When expanding ${\cal L}^{(2)}$ to terms quadratic in the pion field, one finds
\begin{equation}
{\cal L}^{(2)} = (m_u + m_d) f^2 B_0 + \frac{1}{2} \partial_{\mu} \pi_a
\,\partial^{\mu} \pi_a - \frac{1}{2} (m_u + m_d) B_0\, \pi^2_a + 0 (\pi^4) .
\end{equation}
At this point we can immediately verify the GOR relation (26) in
the effective theory. The first (constant) term in (36) corresponds to the
shift of the vacuum energy density by the non-zero quark masses. Identifying
this with the vacuum expectation value of the corresponding piece in the QCD
Lagrangian (1), we find $- (m_u \langle \bar{u} u \rangle + m_d \langle \bar{d}
d \rangle) = (m_u + m_d) f^2 B_0$ and therefore $\langle \bar{u} u \rangle =
\langle \bar{d} d \rangle = - f^2 B_0$ in the chiral limit, $m_{u,d} \to 0$. 
The pion mass term in (36) is evidently identified as $m^2_{\pi} = (m_u + m_d)
B_0$. Inserting $B_0$, we have the GOR relation.

Symmetry breaking terms entering at forth order are of the form
\begin{equation}
\delta{\cal L}^{(4)} = \frac{l_3}{4} B_0^2 \left (Tr [m (U + U^{\dagger})]
\right ) ^2 + \frac{l_4}{4} B_0 \,Tr [\partial_{\mu} U^{\dagger} 
\partial^{\mu} U] \,Tr [m (U + U^{\dagger})] + ...
\end{equation}

e) Given the effective Lagrangian, the framework for systematic perturbative
calculations of the S-matrix involving Goldstone bosons, named Chiral
Perturbation Theory (ChPT), is then defined by the following rules:\\ 
Collect all Feynman diagrams generated by ${\cal L}_{eff}$. Classify
all terms according to powers of a variable $Q$
which stands generically for three-momentum or energy 
of the Goldstone bosons, or for the pion mass $m_{\pi}$. The small
expansion parameter is $Q/4\pi f_{\pi}$. Loops are subject
to dimensional regularisation and renormalisation. 
 
\subsection{Pion-pion scattering}

When using the Gell-Mann - Oakes - Renner relation to leading order in
in the quark mass, it is tacitly assumed that the chiral condensate is
large in magnitude and plays the role of an order parameter for
spontaneous chiral symmetry breaking. This basic scenario has been challenged
(Knecht {\it et al.} 1996) and needs to be 
confirmed. It can in fact be tested by a detailed quantitative analysis
of pion-pion scattering (Leutwyler 2002), the process most extensively studied 
using ChPT.

Consider s-wave $\pi\pi$ scattering at very low energy. The symmetry of
the $\pi\pi$ wave function with $l = 0$ requires isospin to be even,
$I = 0,2$. The corresponding scattering lengths, to leading chiral order,
are
\begin{equation}
a_0^{I=0} = \frac{7\pi}{2}\zeta~~~~,~~~~a_0^{I=2} = -\pi\zeta~~,
\end{equation}
where $\zeta = \left (m_{\pi}/4\pi f_{\pi}\right)^2 \approx 1.45\cdot 10^{-2}$
is the ``small parameter''. Note that the $\pi\pi$ interaction properly 
vanishes in the chiral limit, $m_{\pi}\rightarrow 0$. The next-to-leading
order introduces one-loop iterations of the leading ${\cal L}^{(2)}$ part
of the effective Lagrangian as well as pieces generated by ${\cal L}^{(4)}$.
At that level enters the renormalised coupling constant $\bar{l}_3$ which
also determines the correction to the leading-order GOR relation:
\begin{equation}
m_{\pi}^2~ =~\stackrel{o}{m}_{\pi}^2 -~\frac{1}{2}
\frac{\bar{l}_3}{(4\pi f)^2}
 \stackrel{o}{m}_{\pi}^4 + ~{\cal O}(\stackrel{o}{m}_{\pi}^6)~~,
\end{equation}
where
\begin{equation}
\stackrel{o}{m}_{\pi}^2~=~ - \frac{m_u + m_d}{f^2}\langle \bar{q} q \rangle
\end{equation}
involves the pion mass to leading order in the quark mass and $f$ is the
pion decay constant in the chiral limit. 
An accurate determination of the $I = 0$ s-wave $\pi\pi$ scattering
length therefore provides a constraint for $\bar{l}_3$ which in turn
sets a limit for the next-to-leading order correction to the GOR relation.

Such an investigation has recently been performed (Colangelo {\it et al.} 2001), based on 
low-energy $\pi\pi$ phase shifts extracted from the detailed 
final state analysis of the $K \rightarrow \pi\pi + e\nu$ decay. The result
is that the correction to the leading order prediction (38) is indeed very
small. When translated into a statement about the non-leading term 
entering Eq.(39), it implies that the difference between
$m_{\pi}^2$ and the leading GOR expression (40) is less
than 5 percent. Hence the ``strong condensate'' scenario of spontaneous chiral
symmetry breaking in QCD appears to be confirmed. One should note, however,
that this conclusion is drawn at the level of QCD with only $N_f = 2$
flavours. Additional corrections may still arise when strange quarks are
taken into account. It is nontheless interesting to note that the leading-order 
relationship $m_\pi^2 \sim m_q$ is also 
observed in lattice QCD (Aoki {\it et al.} 2003) up to surprisingly large quark masses.  

\section{The nucleon in the QCD vacuum}

As the lowest-mass excitation of the QCD vacuum with one unit of baryon number,
the nucleon is of fundamental importance to our understanding  of the strong interaction
(Thomas and Weise 2001).
The prominent role played by the pion as a Goldstone boson of spontaneously
broken chiral symmetry has its impact on the low-energy structure
and dynamics of nucleons as well. Decades of research in nuclear physics
have established the pion as the mediator of the long-range force between
nucleons. When probing the individual nucleon itself with long-wavelength
electroweak fields, a substantial part of the response comes from the pion
cloud, the ``soft'' surface of the nucleon. 

The calculational framework for this, baryon chiral perturbation theory, 
has been applied quite successfully to a variety of low-energy processes
(such as threshold pion photo- and electroproduction and Compton scattering on 
the nucleon) for which increasingly accurate experimental data have become
available in recent years. A detailed review can be found in Bernard {\it et al.} (1995). 
An introductory survey is given in Thomas and Weise (2001). 

\subsection{Effective Lagrangian including baryons} 

Let us move to the sector with baryon number $B = 1$ and concentrate
on the physics of the pion-nucleon system, restricting ourselves first to the
case of $N_f = 2$ flavours. (The generalisation to $N_f = 3$ will follow later). 
 
The nucleon is represented by a Dirac spinor field $\Psi_N(x) = (p,n)^T$ 
organised as an isospin-$1/2$ doublet of proton and neutron. The free field
Lagrangian
\begin{equation}
{\cal L}_0^N = \bar{\Psi}_N(i\gamma_{\mu}\partial^{\mu} - M_0)\Psi_N
\end{equation}
includes the nucleon mass in the chiral limit, $M_0$. Note that nucleons, 
unlike pions, are supposed to have at least part of their large mass 
because of the strong mean field provided by the quark condensate 
$\langle\bar{\psi}\psi\rangle$. Such a relationship is explicit, for example,
in the Ioffe formula (28).

We can now construct the low-energy effective Lagrangian for pions 
interacting with a nucleon. The previous pure meson Lagrangian ${\cal L}_{eff}$
is replaced by ${\cal L}_{eff}(U,\partial^{\mu} U, \Psi_N, ...)$ which also
includes the mucleon field. The additional term involving the nucleon, denoted
by ${\cal L}_{eff}^N$, is expanded again in powers of derivatives (external
momenta) and quark masses: 
\begin{equation}
{\cal L}_{eff}^N = {\cal L}_{\pi N}^{(1)} + {\cal L}_{\pi N}^{(2)} ~...
\end{equation}
Let us discuss the leading term, ${\cal L}_{\pi N}^{(1)}$. The modifications
as compared to the free nucleon Lagrangian (45) are twofold. First, there is
a replacement of the $\partial^{\mu}$ term by a chiral covariant derivative
which introduces vector current couplings between the pions and the nucleon.
Secondly, there is an axial vector coupling. This structure of the $\pi N$
effective Lagrangian is again dictated by chiral symmetry. We have
\begin{equation}
{\cal L}_{\pi N}^{(1)} =  \bar{\Psi}_N[i\gamma_{\mu}(\partial^{\mu} 
- i{\cal V}^{\mu}) + \gamma_{\mu}\gamma_5\, {\cal A}^{\mu} - M_0]\Psi_N~~,
\end{equation}
with vector and axial vector quantities involving the Goldstone boson (pion)
fields in the form $\xi = \sqrt{U}$:
\begin{eqnarray}
{\cal V}^{\mu} & = & \frac{i}{2}(\xi^{\dagger}\partial^{\mu}\xi +  
\xi\partial^{\mu}\xi^{\dagger}) = -\frac{1}{4f^2}\varepsilon_{abc}\tau_a~\pi_b
~\partial^{\mu}\pi_c + ...~~, \\
{\cal A}^{\mu} & = & \frac{i}{2}(\xi^{\dagger}\partial^{\mu}\xi -  
\xi\partial^{\mu}\xi^{\dagger}) = -\frac{1}{2f^2}\tau_a~
\partial^{\mu}\pi_a + ...~~,
\end{eqnarray}
where the last steps result when expanding ${\cal V}^{\mu}$ and ${\cal A}^{\mu}$ to 
leading order in the pion fields. 

So far, the only parameters that enter are the nucleon mass, $M_0$,
and the pion decay constant, $f$, both taken in the chiral limit and ultimately
connected with a single scale characteristic of non-perturbative QCD and 
spontaneous chiral symmetry breaking.

When adding electroweak interactions to this scheme, one observes an additional
feature which has its microscopic origin in the substructure of the nucleon,
not resolved at the level of the low-energy effective theory. The analysis of
neutron beta decay $(n \rightarrow p e \bar{\nu})$ reveals that the 
$\gamma_{\mu}\gamma_5$ term in (43) is to be multiplied by the axial vector
coupling constant $g_A$, with the empirical value (Particle Data Group 2004) 
\begin{equation}
g_A = 1.270 \pm 0.003~~.
\end{equation}

At next-to-leading order $({\cal L}_{\pi N}^{(2)})$, the symmetry breaking 
quark mass term enters. It has the effect of shifting the nucleon mass from
its value in the chiral limit to the physical one: 
\begin{equation}
M_N = M_0 + \sigma_N~~.
\end{equation}
The sigma term
\begin{equation}
\sigma_N = m_q\frac{\partial M_N}{\partial m_q} =
\langle N | m_q(\bar{u}u + \bar{d}d) |N\rangle
\end{equation}
measures the contribution of the non-vanishing quark mass, $m_q =
\frac{1}{2}(m_u + m_d)$, to the nucleon mass $M_N$. Its empirical value is in the range
$\sigma_N \simeq (45 - 55)$ MeV and has been deduced by a sophisticated 
extrapolation of low-energy  pion-nucleon data
using dispersion relation techniques (Gasser {\it et al.} 1991, Sainio 2002).

Up to this point, the $\pi N$ effective Lagrangian, expanded to second order
in the pion field, has the form
\begin{eqnarray}
{\cal L}_{eff}^{N} & = & \bar{\Psi}_N(i\gamma_{\mu}\partial^{\mu} 
- M_0)\Psi_N - \frac{g_A}{2f_{\pi}} \bar{\Psi}_N\gamma_{\mu}\gamma_5 
\mbox{\boldmath $\tau$}\Psi_N\cdot\partial^{\mu}\mbox{\boldmath $\pi$}  \\
                   &   & -\frac{1}{4f_{\pi}^2}
 \bar{\Psi}_N\gamma_{\mu} 
\mbox{\boldmath $\tau$}\Psi_N\cdot\mbox{\boldmath $\pi$}\times
\partial^{\mu}\mbox{\boldmath $\pi$}
- \sigma_N\,\bar{\Psi}_N\Psi_N\left(1-{\mbox{\boldmath $\pi$}^2\over 2f_\pi^2}\right) 
+ ...~~,\nonumber
\end{eqnarray}
where we have not shown a series of additional terms of order 
$(\partial^{\mu} \pi)^2$ included in the complete ${\cal L}_{\pi N}^{(2)}$.
These terms come with further constants that need to be fitted to experimental
data.

\subsection{Scalar form factor of the nucleon}

 While pions are well established constituents seen in the electromagnetic structure of the nucleon, its scalar-isoscalar meson cloud is less familiar. On the other hand, the scalar field of the nucleon is at the origin of the intermediate range nucleon-nucleon force, the source of attraction that binds nuclei. Let us therefore have a closer look, guided by chiral effective field theory.

Consider the nucleon form factor related to the scalar-isoscalar quark density, $G_S(q^2) = \langle N(p')|\bar{u}u +  \bar{d}d|N(p)\rangle$, at squared momentum transfer $q^2 = (p - p')^2$. In fact, a better quantity to work with is the form factor $\sigma_N(q^2) = m_qG_S(q^2)$ associated with the scale invariant object $m_q(\bar{u}u +  \bar{d}d)$.  Assume that this form factor can be written as a subtracted dispersion relation:
\begin{equation}
\sigma_N(q^2 = -Q^2) = \sigma_N - {Q^2\over\pi}\int_{4m_\pi^2}^\infty dt{\eta_S(t)\over t(t+Q^2)}~~,
\end{equation}
where the sigma term $\sigma_N$ introduced previously enters as a subtraction constant. We are interested in spacelike momentum transfers with $Q^2 = -q^2 \geq 0$. The dispersion integral in Eq.(50) starts out at the two-pion threshold. It involves the spectral function $\eta_S(t)$ which includes all $J^\pi = 0^+, I=0$ excitations coupled to the nucleon: a continuum of even numbers of pions added to and interacting with the nucleon core. 

Chiral perturbation theory at next-to-next-to-leading order (NNLO) in two-loop approximation has been applied in a recent calculation of the spectral function $\eta_S(t)$ (Kaiser 2003). This calculation includes not only nucleon Born terms and leading $\pi\pi$ interactions but also important effects of intermediate $\Delta(1230)$ isobar excitations in the two-pion dressing of the nucleon core (see Fig.6a). The result (Fig.6b) can be compared with the "empirical" scalar-isoscalar spectral function deduced by analytic continuation from $\pi N$, $\pi\pi$ and $\bar{N}N\leftrightarrow \pi\pi$ amplitudes (H\"ohler 1983). Note that there is no such thing as a "sigma meson" in this spectrum which is completely determined by the (chiral) dynamics of the interacting $\pi\pi$ and $\pi N$ system. \\
\begin{minipage}[t]{6cm}
\centerline {
\includegraphics[width=5cm]{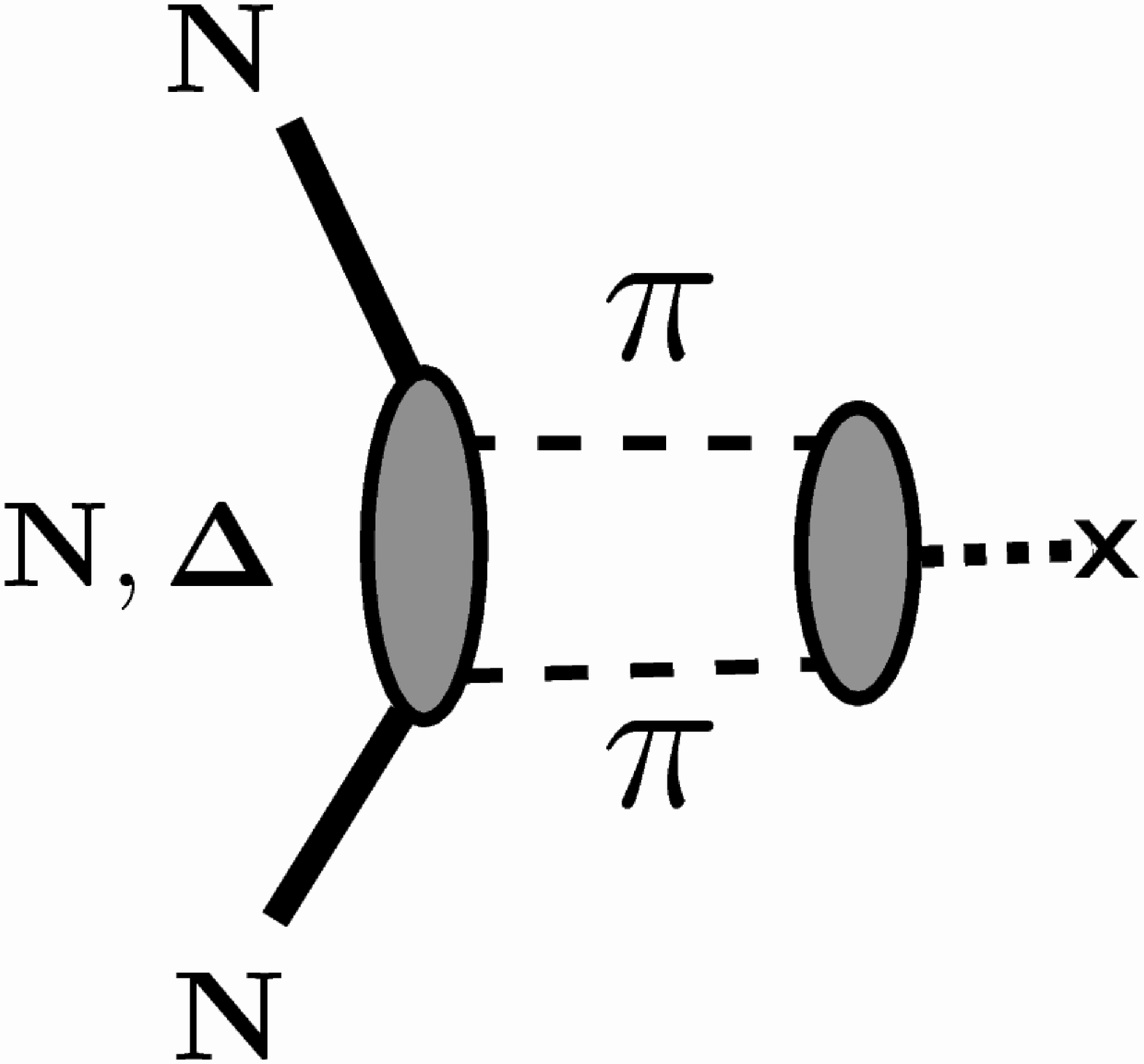}}
\end{minipage}
\hspace{\fill}
\begin{minipage}[t]{6.5cm}
\centerline {
\includegraphics[width=6.5cm]{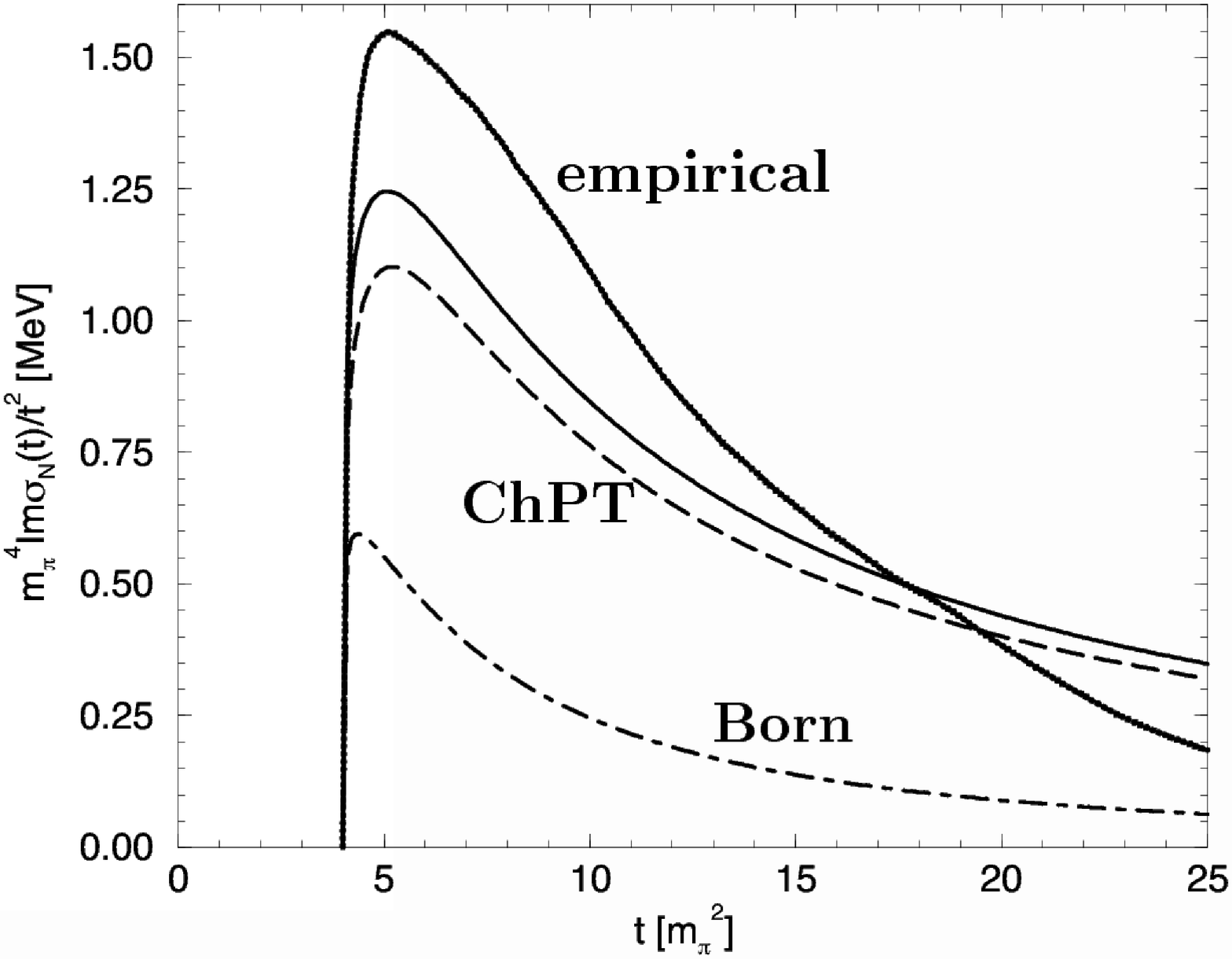}}
\end{minipage}
Figure 6: Left: illustration of the scalar formfactor of the nucleon. Right: spectral function $\eta(t)/t^2$ of the scalar formfactor (50). The empirical spectral distribution is taken from H\"ohler (1983). Also shown are two-loop NNLO chiral perturbation theory (ChPT) calculations (Kaiser 2003). The contribution from the nucleon Born term is given separately.\\
\\
The integral over $\eta_S(t) t^{-2}$ is proportional to the mean squared scalar radius of the nucleon. One finds (Gasser {\it et al.} 1991)
\begin{equation}
\langle r_S\rangle^{1/2} \simeq 1.3\,fm~,
\end{equation}
the largest of all nucleon radii, considerably larger than the proton charge radius of 0.86 fm.  

By its magnitude and range, the form factor $G_S(q^2)$ implies that the nucleon, surrounded with its two-pion cloud, is the source of a strong scalar-isoscalar field with a large effective coupling constant $g_S = G_S(q^2 = 0) = \sigma_N / m_q \simeq 10$. When a second nucleon couples to this scalar field, the resulting two-pion exchange $NN$ interaction $V_{2\pi}$ is reminiscent of a Van der Waals force. More than half of the strength of $V_{2\pi}$ is actually governed by the large spin-isospin polarisability of the nucleon related to the transition $N\rightarrow \Delta$ in the intermediate state. At long and intermediate distances it behaves as (Kaiser {\it et al.} 1998)
\begin{equation}
V_{2\pi}(r) \sim {e^{-2m_\pi r}\over r^6}P(m_\pi r)~ ,
\end{equation}
where $P$ is a polynomial in $m_\pi r$. In the chiral limit $(m_\pi \rightarrow 0)$, this $V_{2\pi}$ approaches the characteristic $r^{-6}$ dependence of a non-relativistic Van der Waals potential. 

The two-pion exchange force is the major source of intermediate range attraction that binds nuclei. This is, of course, not a new observation. For example, the important role of the second-order tensor force from iterated
pion exchange had been emphasised long ago (Brown 1971), as well as the close connection of the nuclear force with the strong spin-isospin polarisability of the nucleon (Ericson and Figureau 1981). The new element that has entered the discussion more recently is the systematics provided by chiral effective field theory in dealing with these phenomena. In fact, in-medium chiral perturbation theory combined with constraints from QCD sum rules is now considered a promising basis for approaching the nuclear many-body problem and finite nuclei in the context of low-energy QCD
(Kaiser {\it et al.} 2002, Finelli {\it et al.} 2004, Vretenar and Weise 2004, Weise 2004).

\subsection{Nucleon mass and pion cloud}

Understanding the nucleon mass is clearly one of the most fundamental issues in nuclear and particle physics (Thomas and Weise 2001).  Progress is now being made towards a synthesis of lattice QCD and chiral effective field theory, such that extrapolations of lattice results to actual observables are beginning to be feasible (Leinweber {\it et al.} 2004, Procura {\it et al.} 2004). Accurate computations of the nucleon mass on the lattice have become available (Ali Khan {\it et al.} 2002, 2004; Aoki {\it et al.} 2003), but so far with $u$ and $d$ quark masses exceeding their commonly accepted small values by typically an order of magnitude. Methods based on chiral effective theory can then be used, within limits, to interpolate between lattice results and physical observables.

The nucleon mass is determined by the expectation value $\langle N| \Theta_\mu^\mu |N\rangle$ of the trace of the QCD energy-momentum tensor (see e.g. Donoghue {\it et al.} 1992),
\begin{eqnarray}
\Theta_\mu^\mu = {\beta(g)\over 2g} G_{\mu\nu}^jG^{\mu\nu}_j + m_u\bar{u}u + m_d\,\bar{d}d + ...\,\,,
\nonumber
\end{eqnarray}
 where $G^{\mu\nu}$ is the gluonic field tensor, $\beta(g) = {\partial g\over \partial \,ln \,\mu}$ is the beta function of QCD, and $m_q\,\bar{q}q$ with $q = u,d,\,$ ... are the quark mass terms (omitting here the anomalous dimension of the mass operator for brevity). Neglecting small contributions from heavy quarks, the nucleon mass taken in the $N_f = 2$ chiral limit, $m_{u,d}\rightarrow 0$, is
\begin{equation}
M_0 = \langle N|{\beta\over 2g}G_{\mu\nu}^jG^{\mu\nu}_j|N\rangle ~.
\end{equation}
This relation emphasises the gluonic origin of the bulk part of $M_N$, the part for which lattice QCD provides an approriate tool to generate the relevant gluon field configurations. At the same time, QCD sum rules connect $M_0$ to the chiral condensate $\langle \bar{q}q\rangle$ as in Ioffe's formula (28). 

In chiral effective field theory, the quark mass dependence of $M_N$ translates into a dependence on the pion mass, $m_\pi^2 \sim m_q$, at leading order. The dressing of the nucleon with its pion cloud, at one-loop order, is illustrated in Fig.7 (left). The systematic chiral expansion of the nucleon mass gives an expression of the form (Procura {\it et al.} 2004):
\begin{equation}
M_N = M_0 +  c m_\pi^2 + d m_\pi^4 - {3\pi \over 2} g_A^2 m_\pi\left({m_\pi \over 4\pi f_\pi}\right)^2 \left(1-{m_\pi^2 \over 8M_0^2}\right) + {\cal O}(m_\pi^6)~,
\end{equation}
where the coefficients $c$ and $d$ multiplying even powers of the pion mass include low-energy constants constrained by pion-nucleon scattering. The coefficient $d$ also involves a $\log m_\pi$ term.
Note that the piece of order $m_\pi^3$ (non-analytic in the quark mass) is given model-independently in terms of the known weak decay constants $g_A$ and $f_\pi$ (strictly speaking: by their values in the chiral limit). The interpolation shown in Fig.7 determines the nucleon mass $M_0$ in the chiral limit and the sigma term $\sigma_N$. One finds $M_0 \simeq 0.89$ GeV and $\sigma_N = (49 \pm 4)$ MeV (Procura {\it et al.} 2004) in this approach.

Admittedly, the gap between presently available lattice results and the actual physical world is still on the large side, perhaps too large for ChPT expansions to extrapolate reliably up to quark masses $m_q > 70$ MeV. However, the synthesis of lattice QCD and chiral effective field theory offers promising perspectives for future developments once lattice QCD begins to operate with quark masses that reach down to pion masses $m_\pi \sim 0.3$ GeV.

\begin{minipage}[t]{4cm}
\centerline {
\includegraphics[width=5cm]{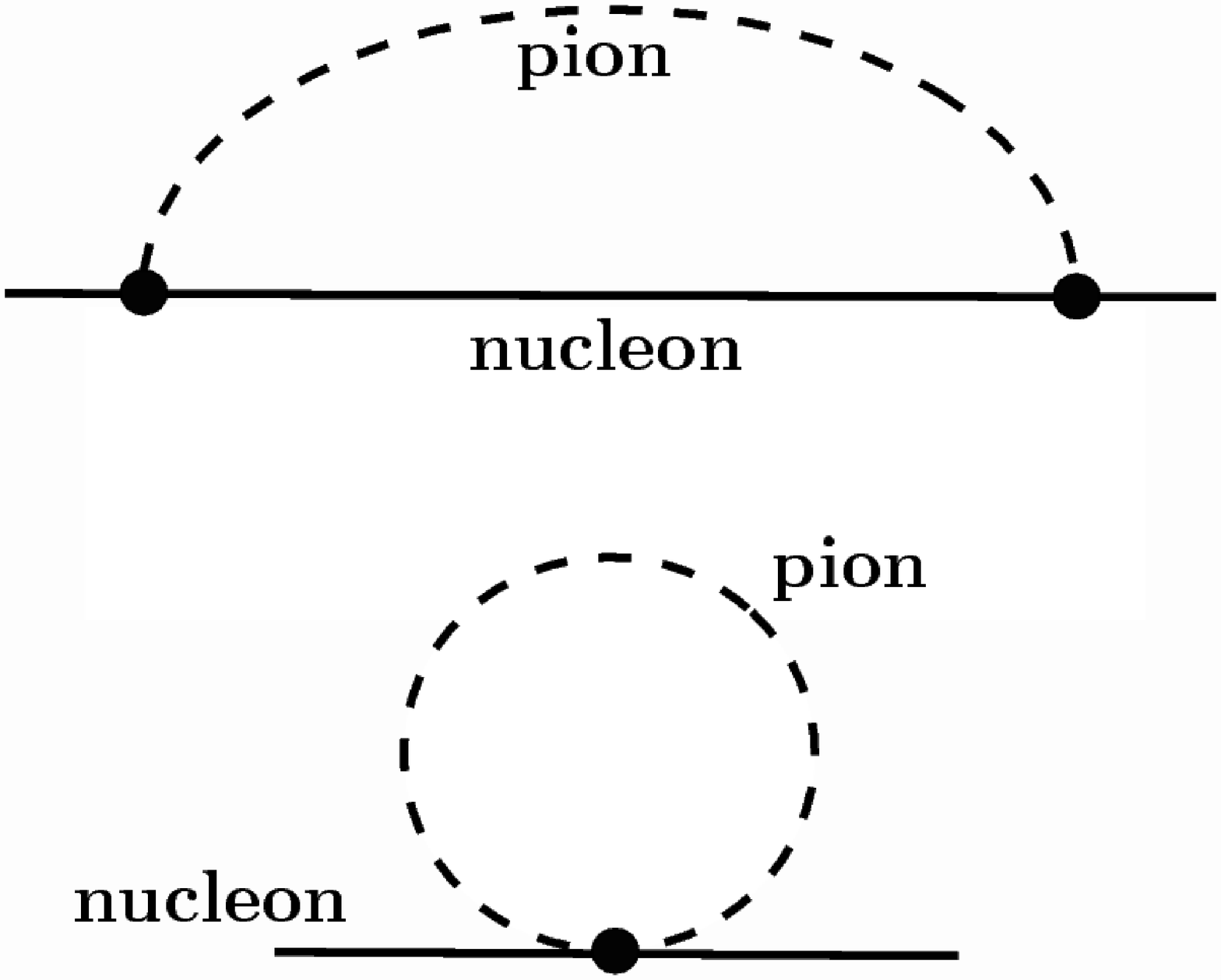}}
\end{minipage}
\hspace{\fill}
\begin{minipage}[t]{8cm}
\centerline {
\includegraphics[width=7.5cm]{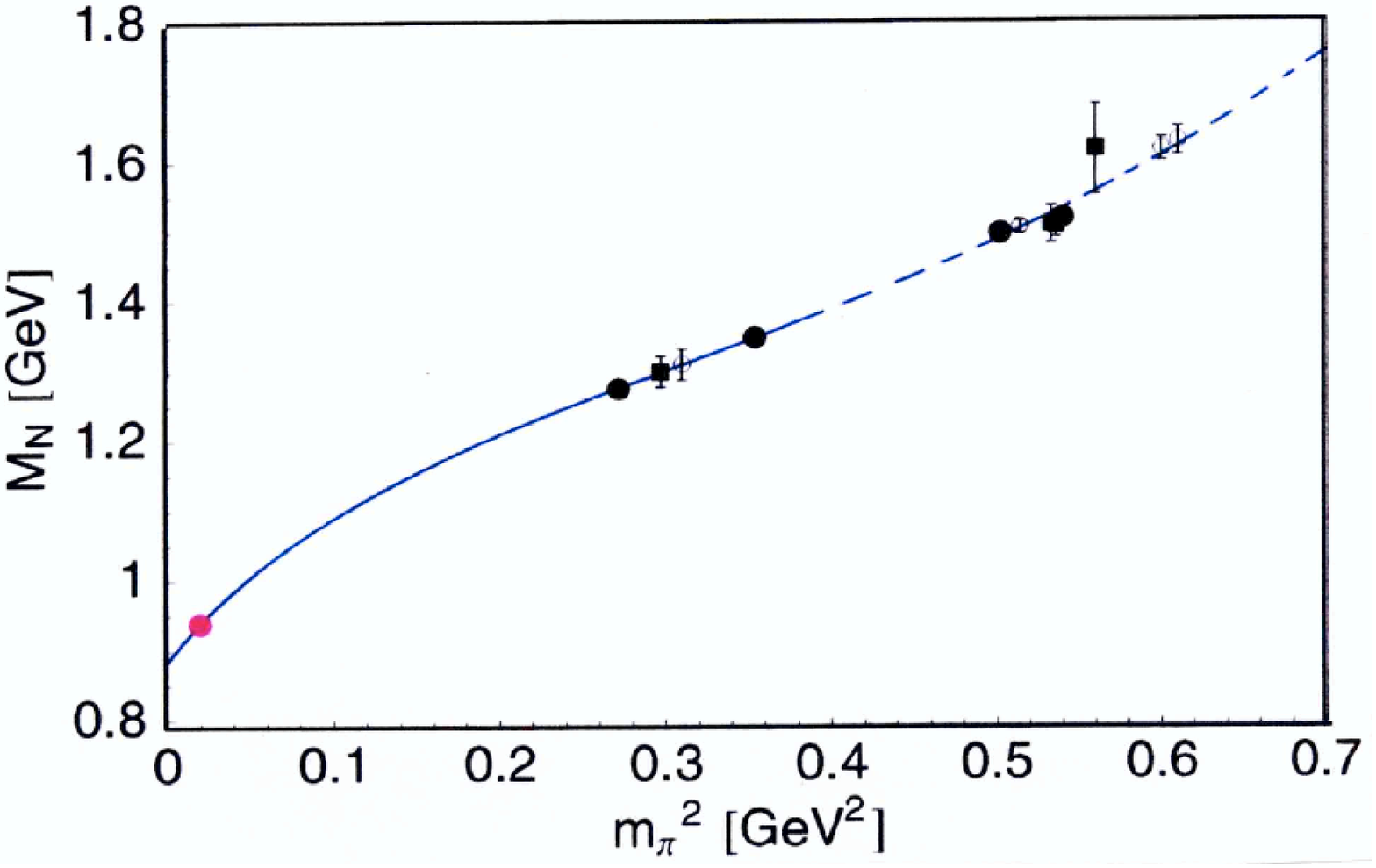}}
\end{minipage}
Figure 7: Left: illustration of pion cloud contributions to the nucleon mass. Right: ChPT interpolation (Procura {\it et al.} 2004) between lattice QCD results for the nucleon mass (at $m_\pi > 0.5$ GeV) and the physical point ($m_\pi = 0.14$ GeV). The lattice QCD data are selected from Ali Khan {\it et al.} (2003) and Aoki {\it et al.} (2003). 

\subsection{Chiral SU(3) dynamics}

Chiral perturbation theory as a systematic expansion in small momenta and quark masses
is limited to low-energy processes with light quarks. It is an interesting issue to what extent 
its generalisation including strangeness can be made to work. The $\bar{K} N$ channel is of particular interest in this context, as a testing ground for chiral SU(3) symmetry in QCD and for the role of explicit chiral symmetry breaking by the relatively large strange quark mass. However, any perturbative approach breaks down in the vicinity of resonances. 
In the $K^- p$ channel, for example, the existence of the $\Lambda(1405)$ resonance 
just below the $K^- p$ threshold renders SU(3) chiral perturbation theory 
inapplicable. At this point the combination with non-perturbative coupled-channels
techniques has proven useful, by generating the $\Lambda(1405)$ dynamically
as an $I=0$ $\bar{K} N$ quasibound state and as a resonance in the $\pi \Sigma$
channel (Kaiser {\it et al.} 1995). This theme actually has its early roots in the coupled-channel K-matrix methods developed in the sixties (Dalitz {\it et al.} 1967). Coupled-channels methods combined with chiral $SU(3)$ dynamics have subsequently been applied to a variety of meson-baryon scattering processes with quite some success (Kaiser {\it et al.} 1997, Oset and Ramos 1998, Caro Ramon {\it et al.} 2000, Oller and Meissner 2001, Lutz and Kolomeitsev 2002). 

{\bf Effective Lagrangian.} The starting point is the chiral $SU(3) \times SU(3)$ effective Lagrangian. The chiral field $U(x)$ of Eq.(30) is now a $3\times 3$ matrix field which collects the full octet of pseudoscalar Goldstone bosons ($\pi, K, \bar{K}, \eta$). The symmetry breaking mass term in (35) is proportional to the quark mass matrix $diag(m_u, m_d, m_s)$ including the mass of the strange quark. The baryon field in (41, 43) is generalized in matrix form, $\Psi_B$, which incorporates the baryon octet ($p, n, \Lambda, \Sigma, \Xi$).
The leading meson-baryon term of the effective Lagrangian which replaces Eq.(49) becomes:
\begin{eqnarray}
{\cal L}_{eff}^B &=&Tr [\bar{\Psi}_B(i\gamma_\mu{\cal D}^\mu - M_0)\Psi_B] \nonumber\\
&+&F\, Tr(\bar{\Psi}_B\gamma_\mu\gamma_5[{\cal A}^\mu,\Psi_B]) + D\,Tr((\bar{\Psi}_B\gamma_\mu\gamma_5{\{\cal A}^\mu,\Psi_B\})
\end{eqnarray}
with the chiral covariant derivative ${\cal D}^\mu\Psi_B = \partial^\mu \Psi_B - i[{\cal V}^\mu,\Psi_B]$. The ${\cal V}^\mu$ and ${\cal A}^\mu$ are the vector and axial vector combinations of octet Goldstone boson
fields which generalise those of Eqs.(44,45). The $SU(3)$ axial vector coupling constants
$D = 0.80\pm 0.01$ and $F = 0.47\pm0.01$ add up to $D + F = g_A = 1.27$. At next-to-leading order, seven additional constants enter in s-wave channels, three of which are constrained by mass splittings in the baryon octet and the remaining four need to be fixed by comparison with low-energy scattering data.

{\bf Coupled channels.} Meson-baryon scattering amplitudes based on the $SU(3)$ effective Lagrangian involve a set of coupled channels. For example, The $K^- p$ system in the isospin $I=0$ sector couples to the $\pi\Sigma$ channel. 
Consider the $T$ matrix ${\bf T}_{\alpha\beta}(p, p'; E)$ connecting meson-baryon channels $\alpha$ and $\beta$ with four-momenta $p, p'$ in the center-of-mass frame:
\begin{equation}
{\bf T}_{\alpha\beta}(p, p') =\\
 {\bf K}_{\alpha\beta}(p, p') + \sum_\gamma\int{d^4q\over (2\pi)^4} {\bf K}_{\alpha\gamma}(p, q)
 \,{\bf G}_\gamma (q)\,{\bf T}_{\gamma\beta}(q, p')\,\, ,
\end{equation}
where ${\bf G}$ is the Green function describing the intermediate meson-baryon loop which is iterated to all orders in the integral equation (56). (Dimensional regularisation with subtraction constants is used in practise). The driving terms ${\bf K}$ in each channel are constructed from the chiral $SU(3)$ meson-baryon effective Lagrangian in next-to-leading order. In the kaon-nucleon channels, for example, the leading terms have the form
\begin{equation}
{\bf K}_{K^\pm p} = 2{\bf K}_{K^\pm n}  = \mp \,2M_N{\sqrt{s} - M_N\over f^2} + ...\,\, ,
\end{equation}
at zero three-momentum, where $\sqrt{s}$ is the invariant c.m. energy and $f$ is the pseudoscalar meson decay constant
($f \simeq 90$ MeV). Scattering amplitudes are related to the $T$ matrix by ${\bf f} = {\bf T}/8\pi\sqrt{s}$.
Note that ${\bf K} > 0$ means attraction, as seen for example in the $K^- p \rightarrow K^- p$ channel.
Similarly, the coupling from $K^- p$ to $\pi\Sigma$ provides attraction, as well as the diagonal matrix elements in the $\pi\Sigma$ channels. Close to the $\bar{K}N$ threshold, we have ${\bf f}(K^- p \rightarrow K^- p)  \simeq m_K/4\pi f^2$, the analogue of the Tomosawa-Weinberg term in pion-nucleon scattering, but now with the (attractive) strength considerably enhanced by the larger kaon mass $m_K$.
 
One should note that when combining chiral effective field theory with the coupled-channels scheme, the "rigorous" chiral counting in powers of small momenta is abandoned in favor of iterating a subclass of loop diagrams to ${\it all}$ orders. However, the gain in physics insights may well compensate for this sacrifice.  Important non-perturbative effects are now included and necessary conditions of unitarity are fulfilled. 

{\bf Low-energy kaon-nucleon interactions}. $K^- p$ threshold data have recently been supplemented by new accurate results for the strong interaction shift and width of kaonic hydrogen (Beer {\it et al.} 2004, Iwasaki {\it et al.} 1997, Ito {\it et al.} 1998). These data, together with existing information on $K^- p$ scattering, the $\pi\Sigma$ mass 
spectrum and measured $K^- p$ threshold decay ratios, set tight constraints on the theory and have therefore revived the interest in this field. Fig.8 shows selected recent results of an improved calculation which combines driving terms from the next-to-leading order chiral $SU(3)$ meson-baryon Lagrangian with coupled-channel equations (Borasoy {\it et al.} 2004). As in previous calculations of such kind, the $\Lambda(1405)$ is generated dynamically, as an $I = 0$ $\bar{K}N$ quasibound state and a resonance in the $\pi\Sigma$ channel. In a quark model picture, this implies that the $\Lambda(1405)$ is not a simple three-quark ($q^3$) state but has a strong $q^4\bar{q}$ component. The detailed threshold behavior of the elastic $K^- p$ amplitude in Fig.8 (right) needs to be further examined in view of the improved accuracy of the most recent kaonic hydrogen data from the DEAR experiment. 
    
\begin{minipage}[t]{5cm}
\centerline {
\includegraphics[width=6cm]{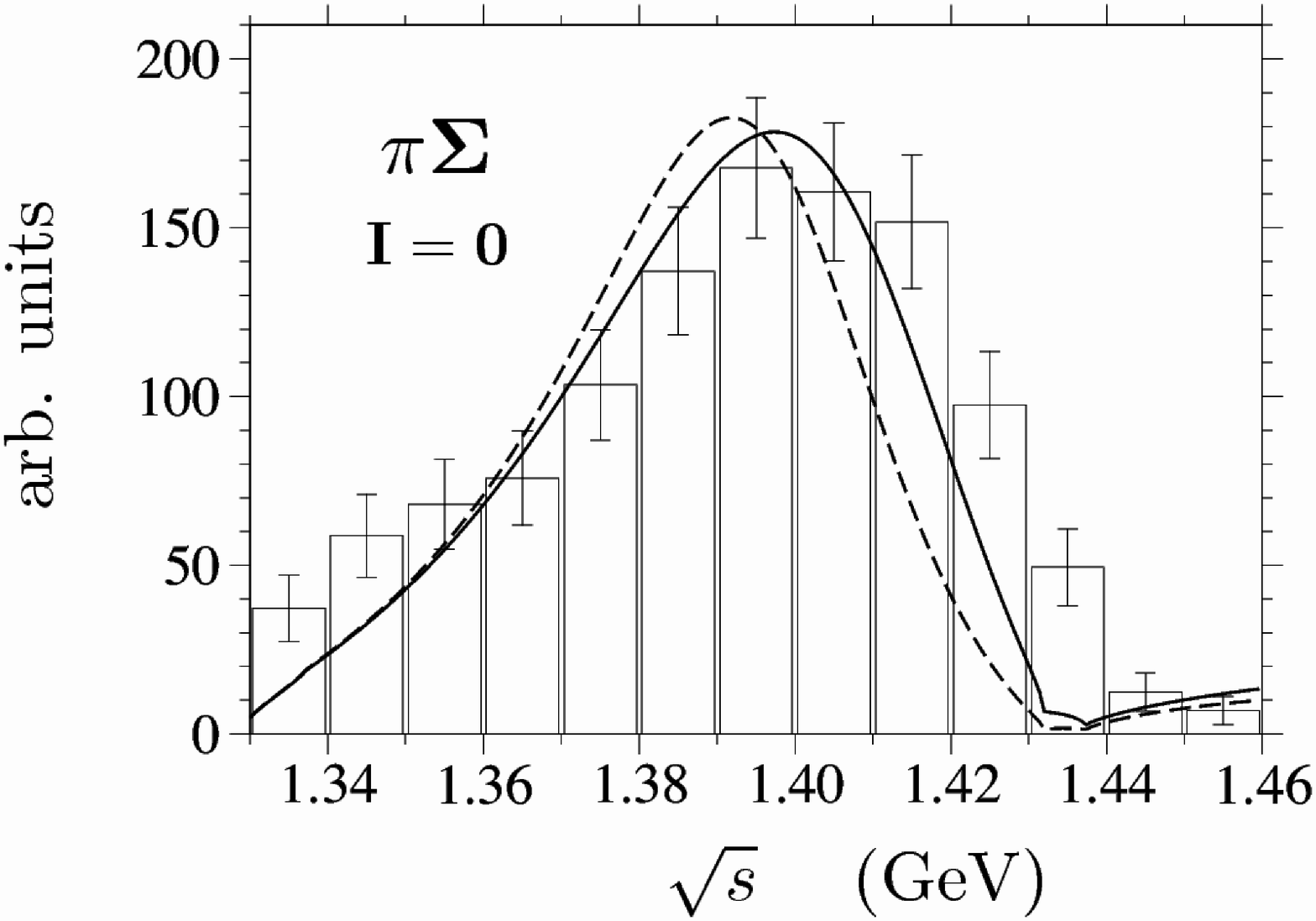}}
\end{minipage}
\hspace{\fill}
\begin{minipage}[t]{6.5cm}
\centerline {
\includegraphics[width=6.5cm]{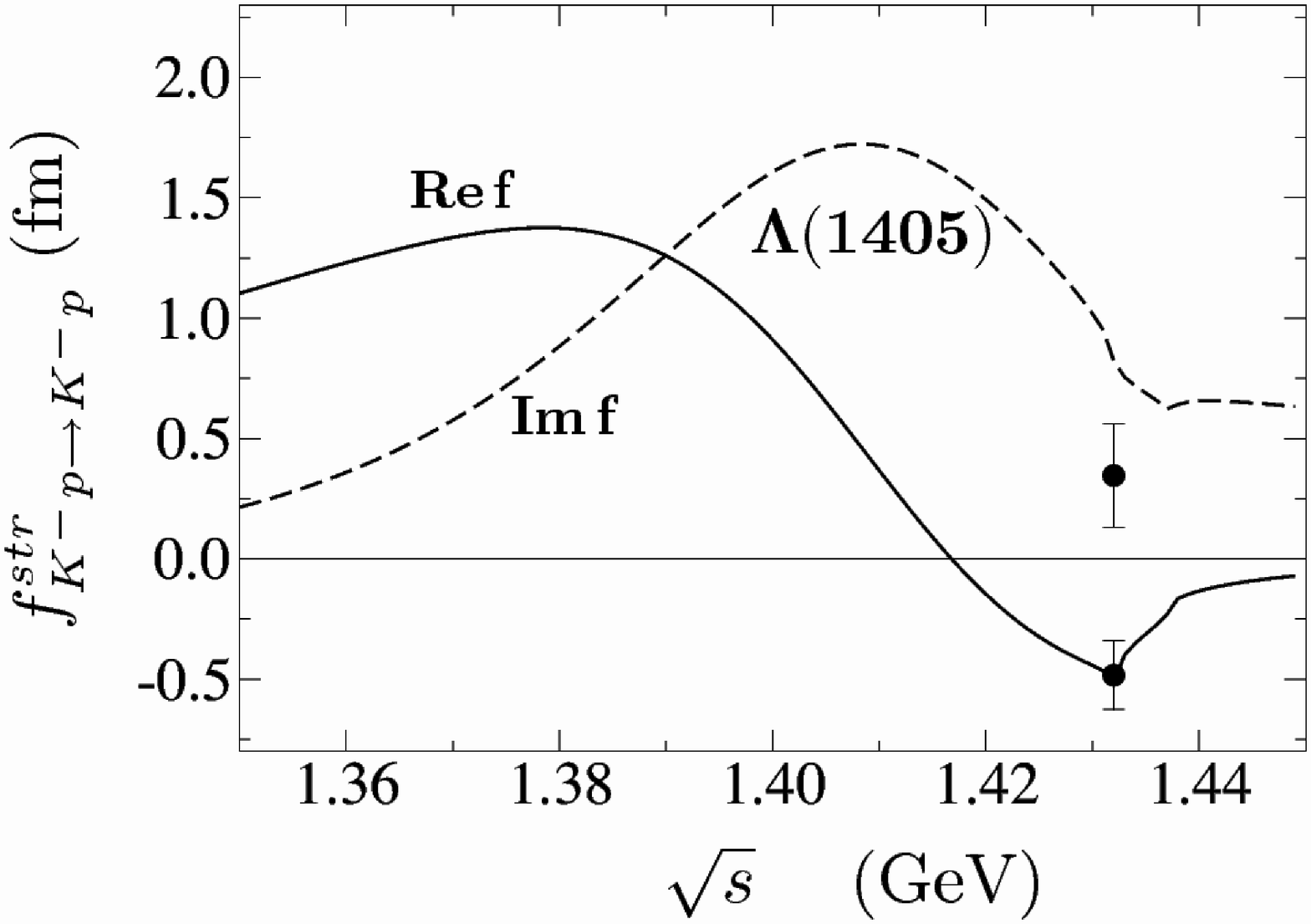}}
\end{minipage}
Figure 8: Left: $\pi\Sigma$ invariant mass spectrum featuring the $\Lambda(1405)$ resonance. Right: real and imaginary parts of the $K^- p$ amplitude. Curves are calculated in the chiral coupled channels approach (Borasoy {\it et al.} 2004). The real and imaginary parts (lower and upper data points) of the $K^- p$ scattering length deduced from the DEAR kaonic hydrogen measurements (Beer {\it et al.} 2004) are also shown. In both figures $\sqrt{s}$ is the invariant $\bar{K}N$ center-of-mass energy.\\
 
\section{Correlations and quasiparticles}

In the preceding chapters our interest has been focused primarily on the low-energy, long-wavelength limit of QCD. In this limit the physics is expressed in terms of color singlet quasiparticle excitations of the QCD vacuum, the lightest hadrons. The spontaneous chiral symmetry breaking scenario in the light-quark sector is the guiding principle for constructing an effective field theory that represents strong interaction physics at momenta small compared to the "chiral gap", $4\pi f_\pi \sim 1$ GeV.

At the other end, on large momentum scales in the multi-GeV range, perturbative QCD operates successfully with quasifree quarks and gluons. Thus there is an obvious and challenging question: how does QCD behave at scales intermediate between those two limits? Do correlations between colored objects exist in such a way that quasiparticle structures are formed which are not observed asymptotically but may have a meaning at intermediate scales? The phenomenological success of the heuristic constituent quark model might indicate such structures. Pairing interactions between constituent quarks as quasiparticles might form new quasiparticles, such as diquarks. The quasiparticle concept has proved highly successful in condensed matter physics. It makes sense to explore its potential relevance also in strongly interacting many-body systems of quarks and gluon fields. Of course, entering this discussion means that one must at least in part rely on phenomenology rather than controlled expansions and approximation schemes. However, lattice QCD results can again provide useful guidance.

\subsection{Gluonic field strength correlations from lattice QCD}

The QCD vacuum hosts not only a condensate of scalar quark-antiquark pairs, but also a strong gluon condensate, the vacuum expectation value of $Tr (G_{\mu\nu}G^{\mu\nu}) \sim {\bf B}^2 -  {\bf E}^2$ where ${\bf B}$ and ${\bf E}$ are the chromomagnetic and chromoelectric fields. We work here in Euclidean space and write
\begin{equation}
{\cal G}_0 \equiv \langle {2\alpha_s\over \pi} Tr \,G_{\mu\nu}\,G_{\mu\nu}\rangle \,\, . 
\end{equation}
with the reduced gluon field tensor $G_{\mu\nu} = G_{\mu\nu}^j\lambda_j/2$. From QCD sum rules and heavy quarkonium spectra one estimates ${\cal G}_0 \approx 1.6$ GeV fm$^{-3}$. The gluon condensate is the local limit $(y\rightarrow x)$ of the gluonic field strength correlation function
(Di Giacomo {\it et al.} 2002):
\begin{equation}
{\cal D}(x,y) \equiv \langle {2\alpha_s\over \pi} Tr \,G_{\mu\nu}(x)\,{\cal S}(x,y)\,G_{\mu\nu}(y)\,{\cal S}^\dagger (x,y)\rangle \sim {\cal G}_0\, exp(-|x-y|/\lambda_g)\,\, ,
\end{equation} 
where the phase ${\cal S}(x,y)$ is a path-ordered exponential of the gauge fields which makes sure that ${\cal D}(x,y)$ is gauge invariant. The non-local behaviour of the correlation function is parametrised by a coherence or correlation length $\lambda_g$.

\centerline {
\includegraphics[width=6cm]{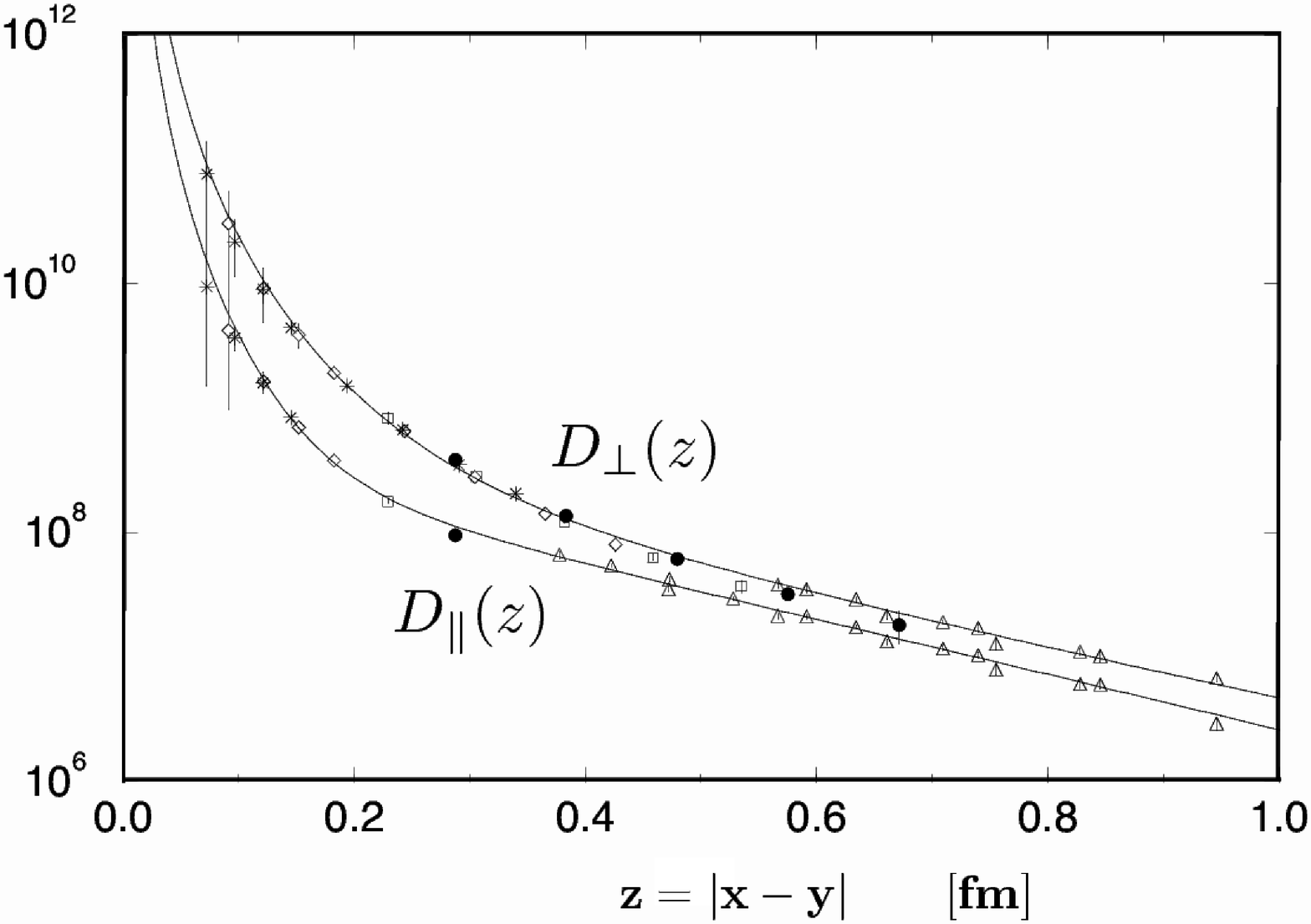}}
\begin{center}
Figure 9: Lattice QCD results of the gluonic field strength correlation function. Taken from Di Giacomo and Panagopoulos (1991) where also the definitions of $D_\parallel$ and $D_\perp$ can be found.
\end{center}

Lattice QCD simulations for the components of ${\cal D}(x,y)$ (Di Giacomo and Panagopoulos 1992) are shown in Fig.9, with the result that the correlation length $\lambda_g$, i.e. the distance over which color fields propagate in the QCD vacuum, is very small:
\begin{equation}
\lambda_g \leq 0.2\, fm\,\, .
\end{equation}
This has interesting consequences. Consider a quark with its (rigorously conserved) color current
\begin{equation}
{\bf J}_\mu^i(x) = \bar{\psi}(x)\gamma_\mu{\lambda^i\over 2}\psi(x) \,\, .
\end{equation}
This current couples to the gluon field. A second quark interacts with the first one by absorbing and emitting gluons whereby they exchange color charges. If the distance over which color can be transported is restricted to the very short range indicated by $\lambda_g$, then the quarks experience an interaction which can be approximated by a local coupling between their color currents:
\begin{equation}
{\cal L}_{int} = -G_c\,{\bf J}_\mu^i(x)\,{\bf J}^\mu_i(x)\,\, ,
\end{equation}
where $G_c \sim \bar{g}^2\, \lambda_g^2$ is an effective coupling strength of dimension $length^2$ which encodes the QCD coupling, averaged over the relevant distance scales, in combination with the squared correlation length, $\lambda_g^2$. 

\subsection{A schematic model: NJL}

This is the starting point of a schematic model that dates back to Nambu and Jona-Lasinio (NJL 1961) and has been further developed and applied to a variety of problems in hadron physics (Vogl and Weise 1991, Hatsuda and Kunihiro 1994). We adopt the non-linear, local interaction (62) and write the following Lagrangian for the quark field $\psi(x)$:
\begin{equation}
{\cal L} = \bar{\psi}(x)(i\gamma^\mu\partial_\mu - m_0)\psi(x) + {\cal L}_{int}(\bar{\psi},\psi)\,\, .
\end{equation}
In essence, by "integrating out" gluon degrees of freedom and absorbing them in the four-fermion interaction ${\cal L}_{int}$, the local $SU(N_c)$ gauge symmetry of QCD is now replaced by a global $SU(N_c)$ symmetry of the NJL model. The mass matrix $m_0$ incorporates small "bare" quark masses. In the limit $m_0 \rightarrow 0$, the Lagrangian (63) evidently has a chiral $SU(N_f) \times SU(N_f)$ symmetry that it shares with the original QCD Lagrangian for $N_f$ massless quark flavors.

A Fierz transform of the color current-current interaction (62) produces a set of exchange terms
acting in quark-antiquark channels. For the $N_f = 2$ case: 
\begin{equation}
{\cal L}_{int} \rightarrow {G\over 2}\left[(\bar{\psi}\psi)^2 + (\bar{\psi}i\gamma_5\vec{\tau}\psi)^2\right] + ... \,\, , 
\end{equation}
where $\vec{\tau} = (\tau_1,\tau_2, \tau_3)$ are the isospin $SU(2)$ Pauli matrices. For brevity we have not shown a series of terms with combinations of vector and axial vector currents, both in color singlet and color octet channels. The constant $G$ is proportional to the color coupling strength $G_c$. Their ratio is uniquely determined by $N_c$ and $N_f$.

{\bf The vacuum sector and constituent quarks.} In the mean field (Hartree) approximation the equation of motion of the NJL model leads to a gap equation
\begin{equation}
M = m_0 - G\langle\bar{\psi}\psi\rangle\,\, ,
\end{equation}
illustrated in Fig.10 (left), which links the dynamical generation of a constituent quark mass $M$ to spontaneous chiral symmetry breaking and the appearance of the quark condensate
\begin{equation}
\langle\bar{\psi}\psi\rangle= -Tr\lim_{\,x\rightarrow \,0^+}\langle {\cal T}\psi(0)\bar{\psi}(x)\rangle = -2iN_fN_c\int {d^4p\over (2\pi)^4}{M\,\theta(\Lambda^2 -\vec{p}\,^2)\over p^2 - M^2 + i\varepsilon}\,\, .
\end{equation}
Starting from $m_0 = 0$ a non-zero mass develops dynamically, together with a non-vanishing chiral condensate, once $G$ exceeds a critical value as shown in Fig.10 (right). The procedure requires a momentum cutoff $\Lambda \simeq 2M$ beyond which the interaction is "turned off". Note that the strong interactions, by polarising the vacuum and turning it self-consistently into a condensate of quark-antiquark pairs, transmute an initially pointlike quark with its small bare mass $m_0$ into a massive quasiparticle with a finite size. Such an NJL-type mechanism is thought to be at the origin of the phenomenological constituent quark masses $M \sim M_N/3$. 

\begin{minipage}[t]{4cm}
\centerline {
\includegraphics[width=5cm]{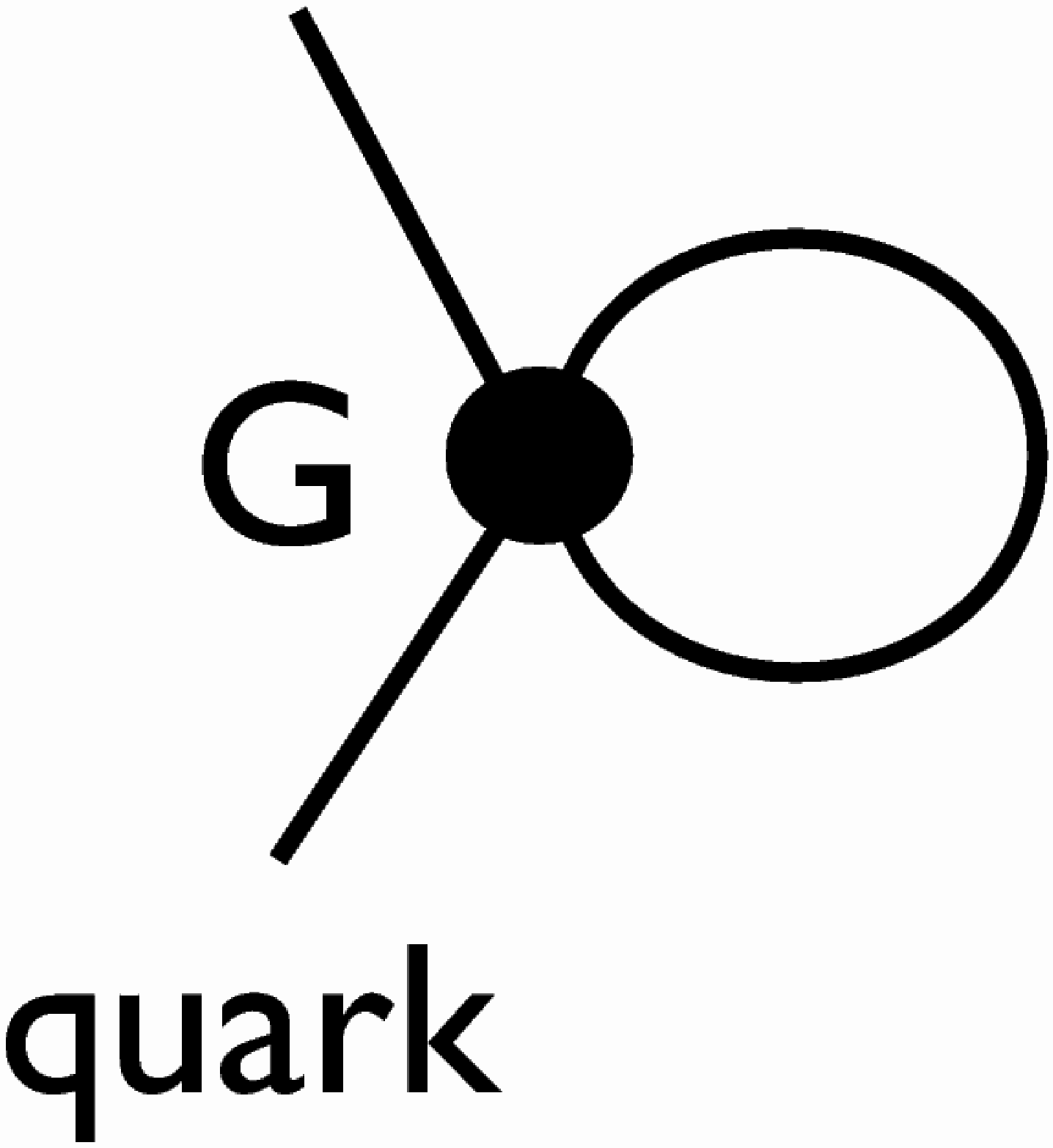}}
\end{minipage}
\hspace{\fill}
\begin{minipage}[t]{7cm}
\centerline {
\includegraphics[width=5.5cm]{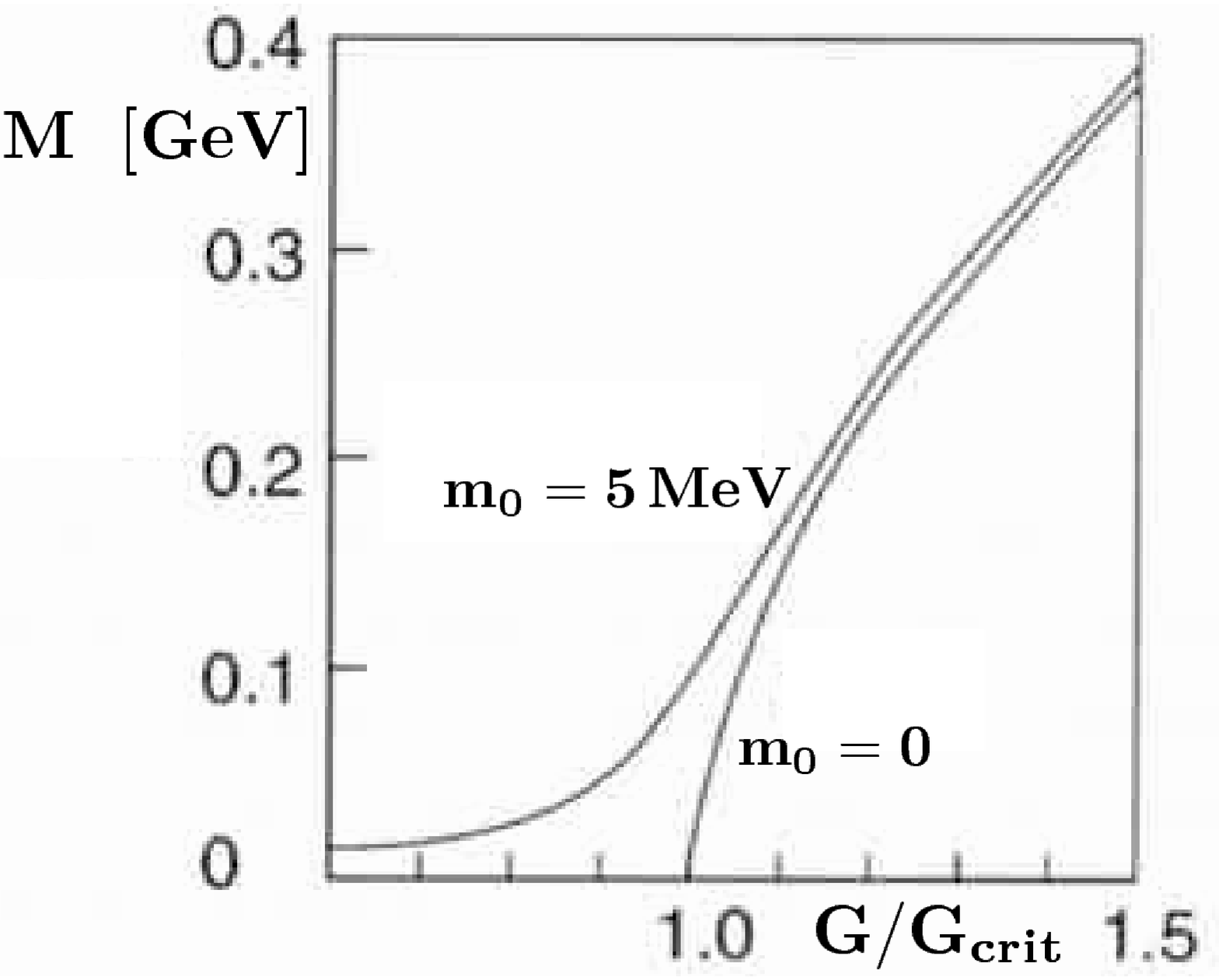}}
\end{minipage}
\begin{center}
Figure 10: Left: illustration of the self-consistent gap equation (65). Right: constituent quark mass as function of the coupling strength G for two values of the bare (current) quark mass $m_0$.\\
\end{center}

Whether this concept of constituent quarks as extended objects inside the proton has observable signatures, is presently an interesting and much discussed issue. By detailed examination of moments of structure functions extracted from inelastic electron-proton scattering at moderate $Q^2$ around 1 GeV$^2$, at energy transfers above the resonance region but below the deep-inelastic (partonic) domain, Petronzio {\it et al.} (2003) have argued that constituent quark sizes of about $0.2 - 0.3$ fm may be visible. However, this effect has to be checked against "standard" QCD evolution down to small $Q^2$.

{\bf The meson sector.} By solving Bethe-Salpeter equations in the color singlet quark-antiquark channels, the NJL model has been used to generate the lightest mesons as quark-antiquark excitations of the correlated QCD ground state with its condensate structure. The framework employed here is analogous to the Random Phase Approximation (RPA) method used frequently to decribe collective particle-hole modes in many-body systems.

We briefly report on earlier results obtained with an NJL model based on the interaction (62) with $N_f = 3$ quark flavors (Klimt {\it et al.} 1990, Vogl and Weise 1991, Hatsuda and Kunihiro 1994). Such a model has a $U(3)_R\times U(3)_L$ symmetry to start with, but the axial $U(1)_A$ anomaly reduces this symmetry to $SU(3)_R\times SU(3)_L\times U(1)_V$. In QCD, instantons are considered responsible for $U(1)_A$ breaking. In the NJL model, these instanton driven interactions are incorporated in the form of a flavor determinant, $\det[\bar{\psi}_a(1 \pm \gamma_5)\psi_b]$ ('t Hooft 1976). This interaction involves all three flavors $u, d, s$ simultaneously in a genuine three-body term. 

The symmetry breaking pattern resulting from such a calculation is demonstrated in the pseudoscalar meson spectrum of Fig.11. Starting from massless $u, d$ and $s$ quarks and in the absence of the 't Hooft determinant, the whole pseudoscalar nonet emerges as a set of massless Goldstone bosons of spontaneously broken $U(3)\times U(3)$. Axial $U(1)_A$ breaking removes the singlet $\eta_0$ from the Goldstone boson sector. Finite quark masses shift the $J^\pi = 0^-$ nonet into its empirically observed position, including $\eta$-$\eta'$ mixing. \\

\centerline {
\includegraphics[width=8cm]{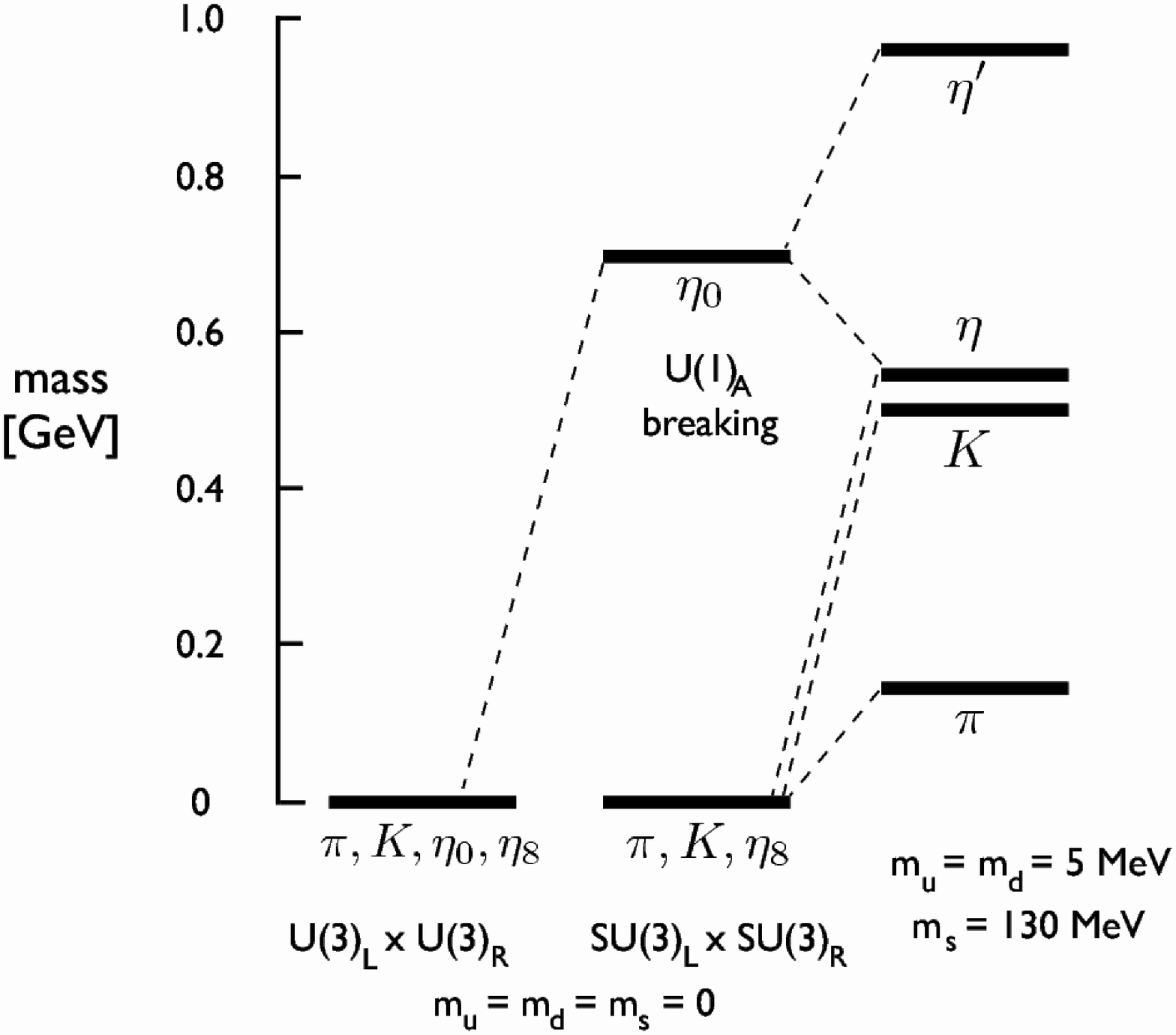}}
Figure 11: Symmetry breaking pattern in the pseudoscalar meson nonet calculated in the $N_f = 3$ NJL model (Klimt {\it et al.} 1990). \\

The combination of self-consistent gap and Bethe-Salpeter equations in the NJL model is an approximation which respects the symmetries of the Lagrangian. The GOR relation (26), 
based on PCAC, is properly fulfilled, and the pion decay constant derives from the basic 
quark-antiquark loop integral as follows:
\begin{equation}
f_\pi^2 = -4i N_c \int{d^4p\over (2\pi)^4} {M^2\,\theta(\Lambda^2 - \vec{p}\,^2)\over (p^2 - M^2 + i\varepsilon)^2}\,\, ,
\end{equation}
with the constituent quark mass $M$ determined self-consistently by the gap equation (65). In practise, the empirical $f_\pi = 92$ MeV is used to fix the momentum cutoff scale $\Lambda$ together with the constituent mass $M_u = M_d = 0.36$ GeV and the chiral condensate $\langle\bar{u}u\rangle = \langle\bar{d}d\rangle = -(250$ MeV$)^3$.

\subsection{Diquarks and pentaquarks}

The one-gluon exchange interaction is attractive for color singlet quark-antiquark pairs and for color antitriplet diquarks. This gives a first hint that attractive correlations in some of those channels might prevail as the gluon-driven interactions develop non-perturbative strength. It is an interesting feature of the NJL model that it establishes close links between quark-antiquark and diquark channels. As in the many-body theory of Fermion systems where particle-particle and particle-hole interactions are connected by crossing relations, the Fierz exchange transformation of the original color current-current interaction (62) has an equivalent representation in diquark channels (Vogl and Weise 1991). For $N_f = 2$ flavors and diquarks interacting in color antitriplet configurations,
\begin{equation}
{\cal L}_{int} \rightarrow 
{H\over 2}\left[(\bar{\psi}_j\,i\gamma_5\,\tau_2\,{\cal C}\,\bar{\psi}_k^T)(\psi_m^T\,{\cal C}^{-1}\,i\gamma_5\,\tau_2\, \psi_n) + \,\, ...\,\,\right]\epsilon^{jkl} \epsilon^{mnl} \,\, ,
\end{equation}
where ${\cal C} = i\gamma_0\gamma_2$ is the charge conjugation operator. Color indices $j, k, l, m, ...\,$ are explicitly displayed. The term written in Eq.(68) applies to scalar diquarks, while a series of additional pieces (representing pseudoscalar, vector and axial vector diquark channels) is not shown explicitly. The coupling strength $H$ is again uniquely determined by the original color coupling $G_c$ via Fierz transformation. 

When solving the Bethe-Salpeter equation for scalar diquarks (Vogl and Weise 1991) in the isoscalar configuration ($ud - du$), one observes strongly attractive correlations which reduce the mass of the diquark as a quasiparticle cluster of two constituent quarks to $M_{ud} \leq 0.3$ GeV, less than half the sum of the constituent masses. Similarly, strong attraction drives the mass of the ($us - su$) scalar diquark far down to $M_{us} \leq 0.6$ GeV from the unperturbed sum of $M_u + M_s \simeq 0.9$ GeV. The strong pairing in spin singlet diquark configurations is a counterpart of the strong attraction observed in pseudoscalar quark-antiquark channels. Other diquark combinations
(pseudoscalar, axial vector etc.) behave differently and stay roughly at the summed mass of their constituent quarks.

QCD at low temperature and very high baryon density is expected to develop Cooper pair condensates and transitions to (color) superconducting phases. Such phenomena are under lively discussion (Alford {\it et al.} 1998, Rapp {\it et al.} 1998). The strongly attractive correlations in scalar diquarks can be considered a precursor of Cooper pairing. \\
\centerline {
\includegraphics[width=5cm]{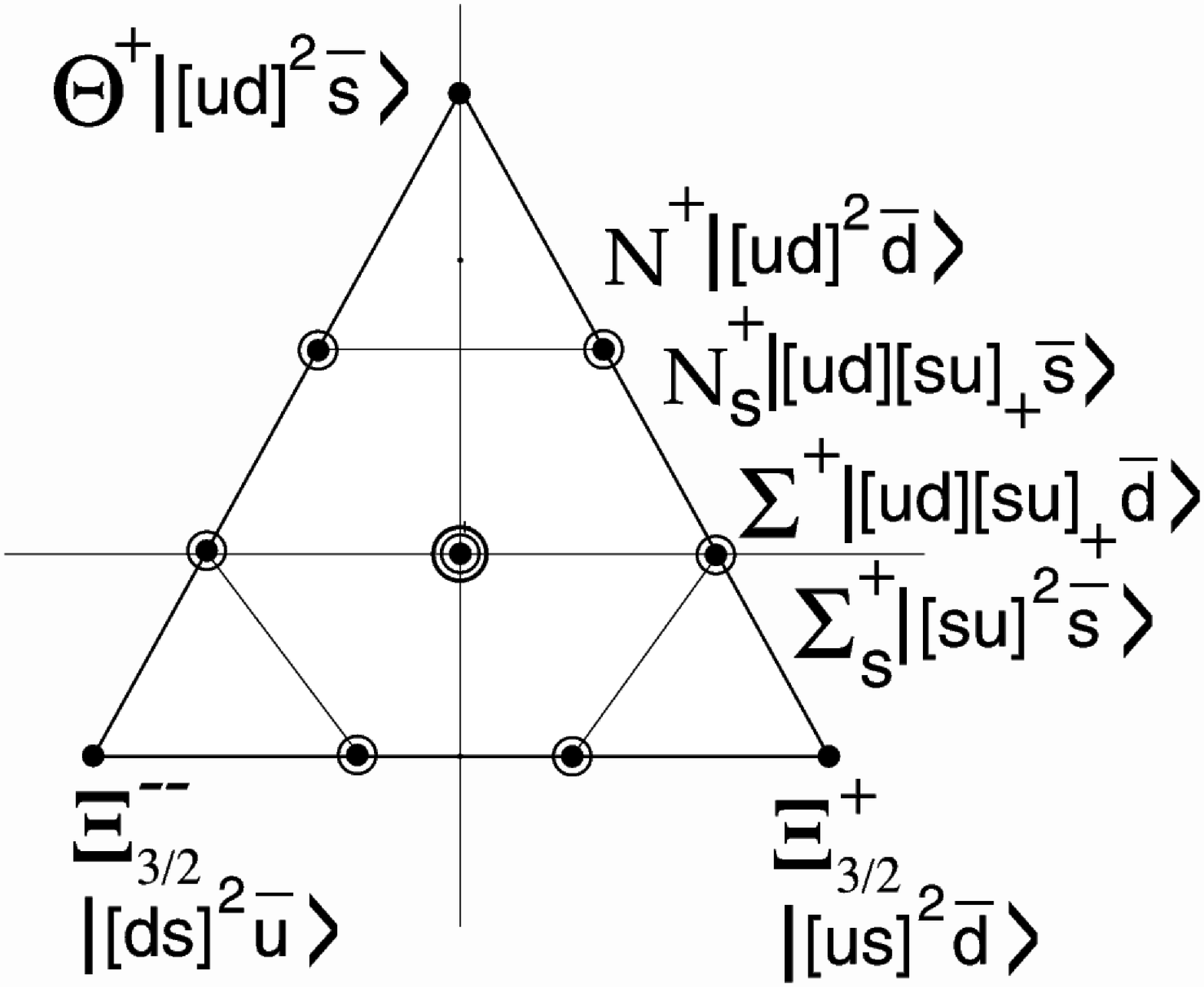}}
Figure 12: Quark content of members of the $(q^4\bar{q})$ antidecuplet + octet representation of flavor $SU(3)$ (from Jaffe and Wilczek 2003).\\

Tightly bound diquark quasiparticles are also at the basis of an interpretation (Jaffe and Wilczek 2003) of the much debated pentaquark state $\Theta^+$, supposedly an "exotic" strangeness $S = +1$ baryon resonance with the quark configuration $(uudd\bar{s})$. Jaffe and Wilczek discuss the $\Theta^+$ as an antidecuplet member of the ${\bf \overline{10} + 8}$ representation of flavor $SU(3)$, as shown in Fig.12. In order for this scheme to work and produce a $\Theta^+$ at the candidate mass around 1540 MeV (see e.g. Close 2004), a $|[ud]^2\bar{s}\rangle$ structure with two scalar $(ud)$ diquarks is anticipated. The required diquark quasiparticle masses in question are indeed within range of those predicted by the previously mentioned NJL model calculation, although it is so far difficult to underscore these statements at a more detailed quantitative level. \\

\subsection{A testing ground: two-colour QCD}

We close these lectures with an outlook into a theoretical laboratory: QCD with $N_c = 2$ colors.
This theory does not have the full complexity of QCD. But it already displays many interesting features of non-abelian gauge theory coupled to spin 1/2 particles. In particular, $N_c = 2$ QCD permits complete lattice simulations of the thermodynamics. Unlike the $N_c = 3$ case in which the Fermion determinant at finite chemical potential becomes complex so that standard Monte Carlo techniques do not work, two-color QCD does not have this problem. The phase diagramme is accessible in the full $(T,\mu)$ plane, at both finite temperature $T$ and baryon chemical potential $\mu$. 

In $N_c = 2$ QCD, diquarks can form color singlets which are the baryons of the theory. The lightest baryons and the lightest quark-antiquark excitations (pions) have a common mass, $m_\pi$, and this spectrum determines the low-energy properties of the theory for small chemical potential. General arguments predict a phase transition to a state with finite baryon density at a critical chemical potential $\mu_c = m_\pi /2$. For $\mu > \mu_c$ Cooper pair condensation of scalar diquarks sets in and the ground state displays superfluidity, associated with spontaneous breaking of the $U(1)_V$ symmetry linked to baryon number.

It is interesting to discuss the thermodynamics of two-color QCD under the aspect of exploring the relevant correlations and identifying the quasiparticles which determine the dynamically active degrees of freedom of such a system. In particular, quasiparticle models of NJL type in which gluons have been integrated out, can be tested in detail against computer simulations of the full theory on large Euclidean lattices. Consider again the NJL Lagrangian  
\begin{equation}
{\cal L} = \bar{\psi}(i\gamma^\mu\partial_\mu - m_0)\psi  - G_c\,\sum_{j = 1}^3\left(\bar{\psi}\gamma^\mu t_j\psi\right)\left(\bar{\psi}\gamma_\mu t_j\psi\right)\,\, ,
\end{equation} 
with a local color current-current coupling between quarks. The $SU(2)_{color}$ generators are denoted $t_j$, and we work in the two-flavor sector with $\psi(x) = (u(x), d(x))^T$. In quark-antiquark and diquark channels, matrix elements of the interaction term of (69) are again conveniently worked out by taking Fierz transforms, as in Eqs.(64, 68). For $N_c = N_f = 2$, the corresponding coupling strengths in
those channels are related to the original coupling $G_c$ of the color currents by $G = H = 3G_c/2$. One notes that there is a symmetry between mesons and diquarks in this model
(a realisation of the so-called Pauli-G\"ursey symmetry).

The partition function is conveniently written by introducing auxiliary boson fields: $\sigma$ and $\phi_a$ for scalar-isoscalar and pseudoscalar-isovector quark-antiquark pairs, respectively; $\Delta$ and $\Delta^*$ for scalar diquarks and antidiquarks.
Rephrasing the Lagrangian (69) in terms of those fields leads to an equivalent Lagrangian 
\begin{eqnarray} 
\tilde{\cal L}& = & \bar{\psi}(i\gamma^\mu\partial_\mu - m_0 + \sigma + i\gamma_5\,\mbox{\boldmath $\tau \cdot \pi$})\psi \nonumber \\ 
& + & {1\over 2}\Delta^*\,\psi^T\,{\cal C}^{-1}\gamma_5\,\tau_2\,t_2\,\psi + 
{1\over 2}\Delta\,\bar{\psi}^T\gamma_5\,\tau_2\,t_2\,{\cal C}\,\bar{\psi} - {\sigma^2 + \mbox{\boldmath$\pi$}^2\over 2G} - {|\Delta|^2\over 2H}\,.
\end{eqnarray} 
The expectation values $\langle\sigma\rangle$ and $\langle|\Delta|\rangle$ represent chiral (quark) and Cooper pair (diquark) condensates. Fixing the constant $G$, the bare quark mass $m_0$ and the momentum space cutoff of the model to reproduce pion properties at $T = \mu = 0$, one can proceed to calculate the thermodynamic potential $\Omega(T, \mu)$ in the mean-field approximation, evaluate the equation of state and derive the phase diagramme. Such calculations (Ratti and Weise 2004) can then be compared with lattice QCD results. At zero chemical potential ($\mu = 0$), Fig.13 shows the characteristic pattern of spontaneous chiral symmetry breaking and restoration as a function of temperature. Above a critical temperature $T_c\simeq 0.2$ GeV, the pion and the scalar ($\sigma$) become degenerate and jointly move away from the low-energy spectrum as a parity doublet, while the pion decay constant $f_\pi$, as order parameter, tends to zero. \\
\centerline {
\includegraphics[width=6cm]{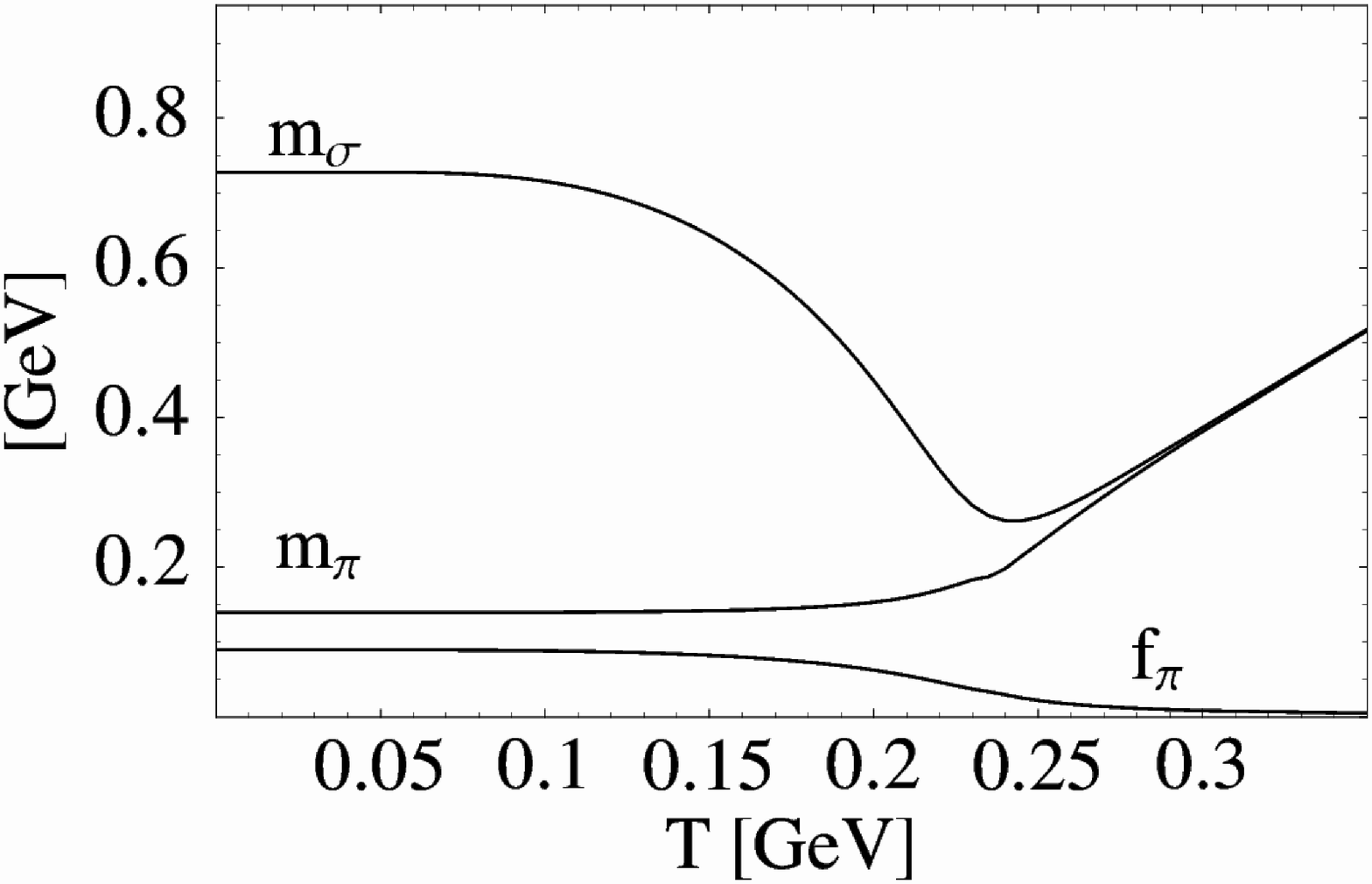}}
Figure 13: Temperature dependence of the pion and scalar masses, and of the pion decay constant, at zero baryon chemical potential (Ratti and Weise 2004).\\

\begin{minipage}[t]{6cm}
\centerline {
\includegraphics[width=5.5cm]{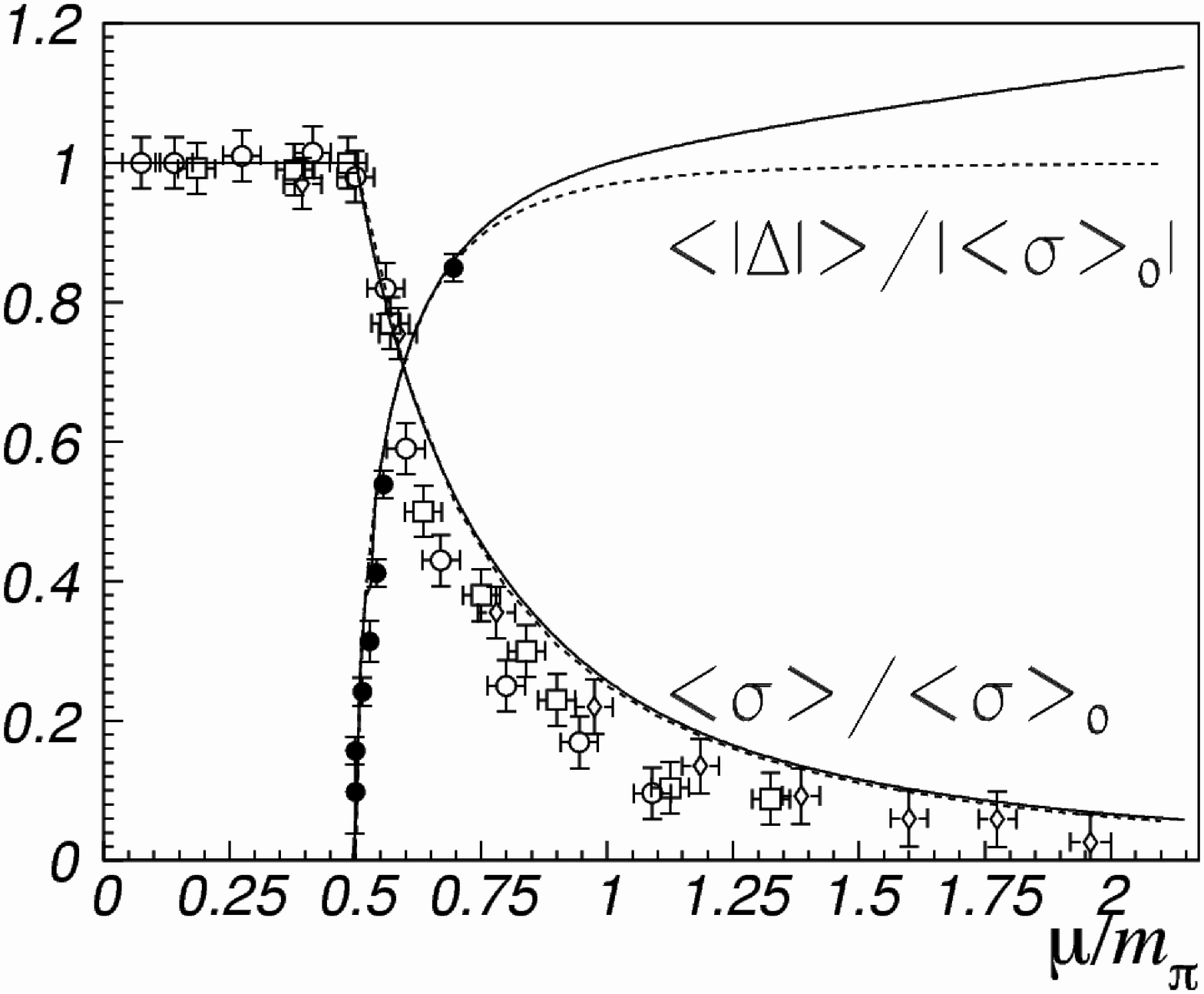}}
\end{minipage}
\hspace{\fill}
\begin{minipage}[t]{6cm}
\centerline {
\includegraphics[width=6cm]{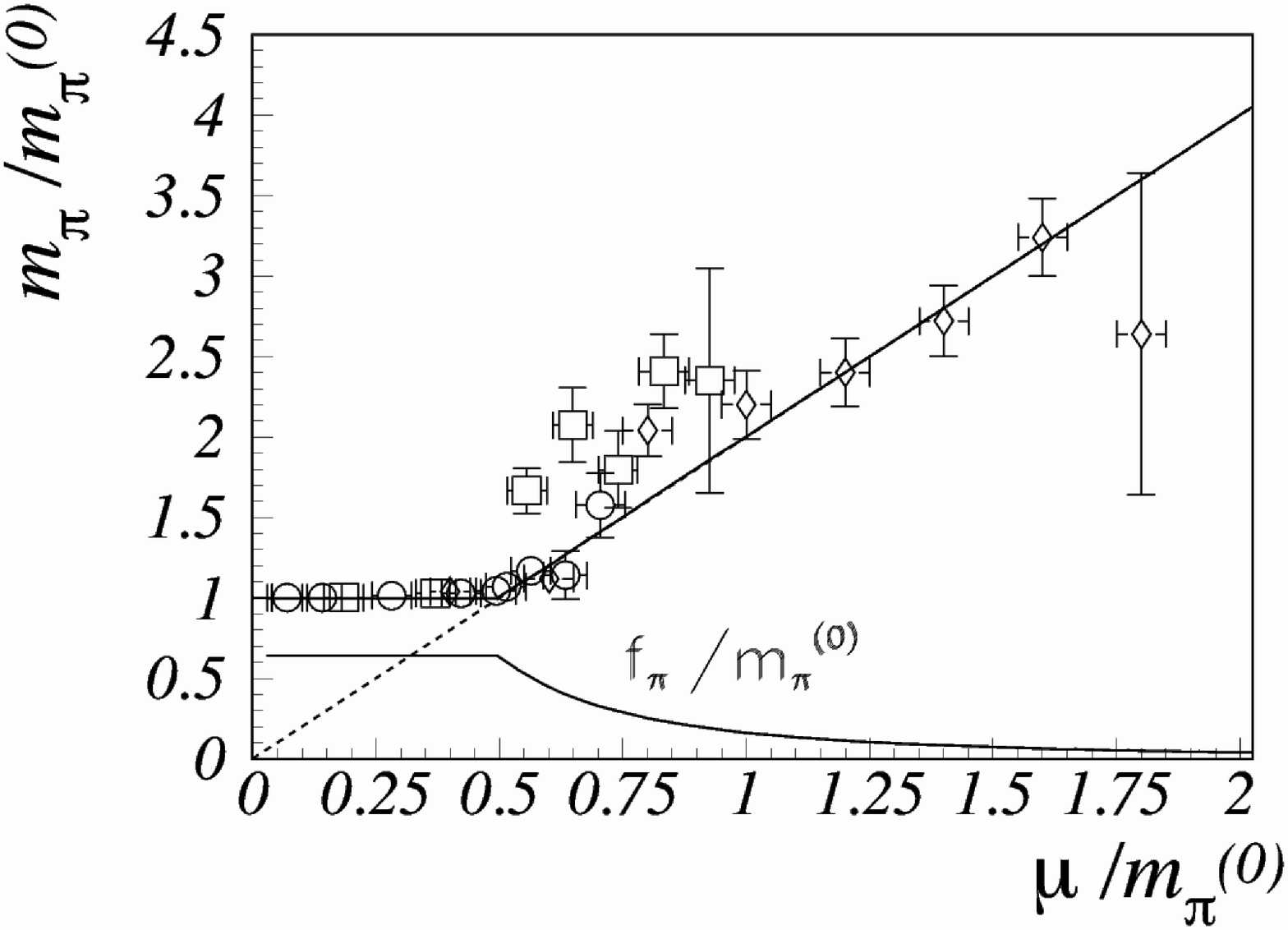}}
\end{minipage}
Figure 14: (Left) chiral and diquark condensates, and (right) pion properties at $T = 0$ as function of baryon chemical potential $\mu$ in $N_c = 2$ lattice QCD (Hands {\it et al.} 2001) and in the NJL model of Eq.(69) (Ratti and Weise 2004). Quantities are given in units of their values at $\mu = 0$.\\

The condensate structure as function of the chemical potential $\mu$ at zero temperature is shown in Fig.14 (left) compared to $N_c = 2$ lattice computations (Hands {\it et al.} 2001). Above a critical chemical potential $\mu_c = m_\pi /2$, the diquark (Cooper pair) condensate develops while at the same time the chiral condensate decreases correspondingly. The pion mass and decay constant follow
this trend as seen in Fig.14 (right).

What is remarkable about these results is that a simple NJL model, with interactions reduced merely to local couplings between the color currrents of quarks, can draw a realistic picture of the quasiparticle dynamics emerging from $N_c = 2$ lattice QCD. One must note that gluon dynamics linked to the original local color gauge symmetry has been replaced by a global color $SU(2)$ symmetry in the model. The symmetry breaking pattern identifies pseudoscalar Goldstone bosons (pions) and scalar diquarks as the thermodynamically active quasiparticles. Color (triplet) quark-antiquark modes which are the remnants of the gluon degrees of freedom in this model, turn out to be far removed from the low-energy spectrum and "frozen". Of course, the low-energy physics of QCD differs qualitatively between $N_c = 2$ and $N_c = 3$ because of the very different nature of the baryonic quasiparticles in these two theories. However, the present example demonstrates that a synthesis of lattice QCD strategies with effective field theories and quasiparticle approaches promises further insights.\\
\\

Special thanks go to my collaborators Bugra Borasoy, Thomas Hemmert,  Norbert Kaiser, Robin Nissler, Massimiliano Procura and Claudia Ratti whose recent works have contributed substantially to the subjects presented in these lectures. Partial support by BMBF, DFG and GSI is gratefully acknowledged.

\section*{References}
\frenchspacing
\begin{small}
\reference{Alford M G, Rajagopal K, and Wilczek F, 1998, {\it Phys. Lett.} {\bf B 422} 247.}
\reference{Ali Khan A {\it et al.}, 2002, {\it Phys. Rev.} {\bf D 65} 054505.}
\reference{Ali Khan A {\it et al.}, 2004, {\it Nucl. Phys.} {\bf B 689} 175.}
\reference{Aoki S {\it et al.}, 2003, {\it Phys. Rev.} {\bf D 68} 054502.}
\reference{Ackerstaff K {\it et al.}, OPAL Collaboration, 1999, {\it Eur. Phys. J.} {\bf C 7} 571.}
\reference{Bali G S, Schilling K, and Wachter A, 1997, {\it Phys. Rev.} {\bf D 56} 2566.}
\reference{Bali G S, 2001, {\it Phys. Reports} {\bf 343} 1.}
\reference{Barate M {\it et al.}, ALEPH Collaboration, 1998, {\it Eur. Phys. J.} {\bf C 4} 409.}
\reference{Bernard V, Kaiser N, and Meissner U-G, 1995, {\it Int. J. Mod. Phys.} {\bf E 4} 193.}
\reference{Borasoy B, Nissler R, and Weise, 2004, hep-ph/0410305.}
\reference{Boyd G {\it et al.}, 1995, {\it Phys. Lett.} {\bf B 349} 170.}
\reference{Brown G E, 1971, {\it Unified Theory of Nuclear Models and Forces}, 3rd ed.,\\ North-Holland, Amsterdam.}
\reference{Caro Ramon J, Kaiser N, Wetzel S, and Weise W, 2000, {\it Nucl. Phys.} {\bf A 672} 249.}
\reference{Close F E, 2004, {\it Lectures on The Quark Model}, these Proceedings.}
\reference{Colangelo G, Gasser J, and Leutwyler H, 2001, {\it Nucl. Phys.} {\bf B 603} 125.}
\reference{Dalitz R H, Wong T-C, and Rajasekaran G, 1967, {\it Phys. Rev.} {\bf 153} 1617.}
\reference{Davies Ch, 2004, {\it Lectures on Lattice QCD}, these Proceedings.}
\reference{Di Giacomo A and Panagopoulos, 1992, {\it Phys. Lett.} {\bf B 285} 133.}
\reference{Di Giacomo A, Dosch H G, Shevchenko V I, and Simonov Y A, 2002, {\it Phys. Reports} {\bf 372} 319.}
\reference{Donoghue J F, Golowich E, and Holstein B R, 1992, {\it Dynamics of the Standard Model}, \\Cambridge University Press.}
\reference{Ericson M and Figureau A, 1981, {\it J. Phys.} {\bf G 7} 1197.}
\reference{Fritzsch H, Gell-Mann M, and Leutwyler H, 1973, {\it Phys. Lett.} {\bf B 47} 365.}
\reference{Finelli P, Kaiser N, Vretenar D, and Weise W, 2004, {\it Nucl. Phys.} {\bf A 735} 449.}
\reference{Gasser J and Leutwyler H, 1984, {\it Ann. of Phys.} {\bf 158} 142.}
\reference{Gasser J, Leutwyler H, and Sainio M E, 1991, {\it Phys. Lett.} {\bf B 253} 252, 260.}
\reference{Gell-Mann M, Oakes R, and Renner B, 1968, {\it Phys. Rev.} {\bf 175} 2195.}
\reference{Goltermann M and Peris S, 2000, {\it Phys. Rev.} {\bf D 61} 034018.}
\reference{Gross D J and Wilczek F, 1973, {\it Phys. Rev. Lett.} {\bf 30} 1343.}
\reference{Hands S, Montvay I, Scorzato L, and Skullerud J, 2001, {\it Eur. Phys. J.} {\bf C 22} 451.}
\reference{Hatsuda T and Kunihiro T, 1994, {\it Phys. Reports} {\bf 247} 221.}
\reference{H\"ohler G, 1983, {\it Pion-Nucleon Scattering}, in: Landolt-B\"ornstein New Series
Vol.I/9b2, \\Springer, Berlin.}
\reference{'t Hooft G, 1976, {\it Phys. Rev.} {\bf D 14} 3432.}
\reference{Jaffe R L and Wilczek F, 2003, {\it Phys. Rev. Lett.} {\bf 91} 232003.}
\reference{Kaiser N, Siegel P B, and Weise W, 1995, {\it Nucl. Phys.} {\bf A 594} 325.}
\reference{Kaiser N, Waas T, and Weise W, 1997, {\it Nucl. Phys.} {\bf A 612} 297.}
\reference{Kaiser N, Gerstend\"orfer S, and Weise W, 1998, {\it Nucl. Phys.} {\bf A 637} 395.}
\reference{Kaiser N, Fritsch S, and Weise W, 2002, {\it Nucl. Phys.} {\bf A 700} 343.}
\reference{Kaiser N, 2003, {\it Phys. Rev.} {\bf C 68} 025202.}
\reference{Karsch F, Laermann E, and Peikert A, 2000, {\it Nucl. Phys.} {\bf B 605} 597.}
\reference{Klimt S, Lutz M, Vogl U, and Weise W, 1990, {\it Nucl. Phys.} {\bf A 516} 429.}
\reference{Knecht M, Moussalam B, Stern J, and Fuchs N H, 1996, {\it Nucl. Phys.} {\bf B 471} 445.}
\reference{Klingl F and Weise W, 1999, {\it Eur. Phys. J.} {\bf A 4} 225.}
\reference{Leinweber D B, Thomas A W, and Young R D, 2004, {\it Phys. Rev. Lett.} {\bf 92} 242002.}
\reference{Leutwyler H, 1994, {\it Ann. of Phys.} {\bf 235} 165.}
\reference{Leutwyler H, 2002, {\it Nucl, Phys. B Proc. Suppl.} {\bf 108} 37.}
\reference{Lutz M F M and Kolomeitsev E, 1999, {\it Nucl. Phys.} {\bf A 700} 193.}
\reference{Marco E and Weise W, 2000, {\it Phys. Lett.} {\bf B 482} 87.}
\reference{Metag V, 2004, {\it Lectures on Hadrons in the Nuclear Medium}, these Proceedings.}
\reference{Nambu Y and Jona-Lasinio G, 1961, {\it Phys. Rev.} {\bf 122} 345.}
\reference{Oller J A and Meissner U-G, 2001, {\it Phys. Lett.} {\bf B 500} 263.}
\reference{Oset E and Ramos A, 2000, {\it Nucl. Phys.} {\bf A 635} 99.}
\reference{Particle Data Group, S. Eidelman {\it et al.}, 2004, {\it Review of Particle Properties},  {\it Phys. Lett. } {\bf B 592} 1.}
\reference{Petronzo R, Simula S, and Ricco G, 2003,  {\it Phys. Rev.} {\bf D 67} 094004.}
\reference{Pich A and Prades J, 2000,  {\it Nucl. Phys. Proc. Suppl.} {\bf 86} 236.}
\reference{Politzer H D, 1973,  {\it Phys. Rev. Lett.} {\bf 30} 1346.}
\reference{Procura M, Hemmert T, and Weise W, 2004,  {\it Phys. Rev.} {\bf D 69} 034505.}
\reference{Rapp R, Sch\"afer T, Shuryak E V, and Velkovsky M, 1998,  {\it Phys. Rev. Lett.} {\bf 81} 53.}
\reference{Ratti C and Weise W, 2004, {\it Phys. Rev.} {\bf D 70} 054013.}
\reference{Sainio M E, 2002,  {\it PiN Newslett.} {\bf 16} 138 (hep-ph/0110413).}
\reference{Thomas A W and Weise W, 2001, {\it The Structure of the Nucleon},  Wiley-VCH, Berlin.}
\reference{Vogl U and Weise W, 1991, {\it Prog. Part. Nucl. Phys.},  {\bf 27}, 195.}
\reference{Vretenar D and Weise W, 2004, {\it Lect. Notes Phys. (Springer)},  {\bf 641}, 65.}
\reference{Weinberg S, 1967, {\it Phys. Rev. Lett.} {\bf 18} 507.}
\reference{Weinberg S, 1979, {\it Physica} {\bf A 96} 327.}
\reference{Weise W, 2003, {\it Chiral Dynamics and the Hadronic Phase of QCD}, in: Proc. Int. School of Physics "Enrico Fermi", Course CLIII (Varenna 2002), eds. A. Molinari {\it et al.}, IOS Press, Amsterdam.}
\reference{Weise W, 2004, {\it Nucl. Phys.} {\bf A}, in print (nucl-th/0412075).}
\end{small}
\nonfrenchspacing

 \end{document}